\documentclass[useAMS,usenatbib]{mn2e}
\usepackage{graphicx,txfonts,amssymb,amsfonts,natbib,times,hyperref}

\begin{document}

\title[Cosmic shear]
  {Cosmic shear statistics in cosmologies with non-Gaussian initial conditions}

\author[C.\,Fedeli \& L. Moscardini]{C.\,
      Fedeli$^{1,2,3}$ and L.\,Moscardini$^{1,2,3}$\\$^1$ 
      Dipartimento di Astronomia, Universit\`a di Bologna,
      Via Ranzani 1, I-40127 Bologna, Italy (cosimo.fedeli@unibo.it)\\$^2$ INAF-Osservatorio
      Astronomico di Bologna, Via Ranzani 1, I-40127 Bologna, Italy\\$^3$
      INFN, Sezione di Bologna, Viale Berti Pichat 6/2, I-40127 Bologna, Italy\\}
      
\maketitle

\begin{abstract}
We computed the power spectrum of weak cosmic shear in models with non-Gaussian primordial density fluctuations. Cosmological initial conditions deviating from Gaussianity have recently attracted much attention in the literature, especially with respect to their effect on the formation of non-linear structures and because of the bounds that they can put on the inflationary epoch. The fully non-linear matter power spectrum was evaluated with the use of the physically motivated, semi-analytic halo model, where the mass function and linear halo bias were suitably corrected for non-Gaussian cosmologies. In agreement with previous work, we found that a level of non-Gaussianity compatible with CMB bounds and with positive skewness produces an increase in power of the order of a few percent at intermediate scales. We then used the matter power spectrum, together with observationally motivated background source redshift distributions in order to compute the cosmological weak lensing power spectrum. We found that the degree of deviation from the power spectrum of the reference Gaussian model is small compared to the statistical error expected from even future weak lensing surveys. However, summing the signal over a large range of multipoles can beat down the noise, bringing to a significant detection of non-Gaussianity at the level of $|f_\mathrm{NL}| \simeq $ few tens, when all other cosmological parameters are held fixed. Finally, we have shown that the constraints on the level of non-Gaussianity can be improved by $\sim 20\%$ with the use of weak lensing tomography. 
\end{abstract}

%%%%%%%%%%%%%%%%%%%%%%%%%%%%%%%%%%%%%%%%%%%%%%%%%%%
\section{Introduction}\label{sct:introduction}
%%%%%%%%%%%%%%%%%%%%%%%%%%%%%%%%%%%%%%%%%%%%%%%%%%%

One of the major success of the inflationary scenario for the early Universe is that it explains the formation of the seed fluctuations in the dark matter density field that, due to gravitational instability, eventually formed non-linear structures such as galaxies, galaxy clusters and voids (\citealt{GU81.1,BR84.1,KO87.1}). In the simplest model of inflation, the early accelerated expansion phase of the Universe was driven by a single, minimally coupled scalar field. In this case, density fluctuations are predicted to follow an almost Gaussian probability distribution. Significant deviations from Gaussianity are however predicted in many of the more elaborated models of inflation that have been developed up to date (\citealt*{MA86.1,AL87.1}).

The most recent analysis of the Cosmic Microwave Background radiation power spectrum of temperature fluctuations (CMB henceforth, \citealt{DU09.1,KO09.1}, see also \citealt*{SM09.1}) is consistent with Gaussian primordial density perturbations, although a significant level of non-Gaussianity is still allowed. Inflationary models exist predicting a scale dependent behavior for the non-Gaussian amplitude \citep{LO08.1}, implying that the amount of deviation from Gaussianity might be different between the large scales probed by the CMB and the small scales probed by galaxies and galaxy clusters.

In light of this, it is important to understand the effect on structure formation in the Universe of non-Gaussianity levels compatible with CMB bounds and/or with a significant scale evolution. This kind of problem has recently attracted much attention in the literature, with efforts directed towards the abundance of nonlinear structures (\citealt*{MA00.2}; \citealt{VE00.1}; \citealt*{MA04.1,KA07.1}; \citealt{GR07.1,GR09.1,MA09.2}), halo biasing (\citealt{DA08.1,MC08.1}; \citealt*{FE09.1}), galaxy bispectrum \citep{SE07.2,JE09.1}, mass density distribution (\citealt{GR08.2}) and topology \citep{MA03.2,HI08.2}, integrated Sachs-Wolfe effect \citep*{AF08.1,CA08.1}, Ly-$\alpha$ flux from low-density intergalactic medium \citep{VI09.1}, $21$ cm fluctuations \citep*{CO06.2,PI07.1} and reionization (\citealt{CR09.1}). 

One particular observable quantity that should be affected in a non-trivial way by non-Gaussianity is the fully non-linear power spectrum of the large-scale dark matter distribution. Studies of the effect of non-Gaussian initial conditions on this observable have been recently put forward with numerical $n$-body simulations \citep{GR08.2}, with renormalized perturbation theory \citep*{TA08.1} and by using both \citep*{GI09.1}. Although differences exist between different works, they all agree in setting the effect of non-Gaussianity to a few percent at most on mildly non-linear scales.

Observationally, the matter power spectrum on scales smaller than CMB scales is usually measured by looking at the distribution of pairs of galaxies, that are known to be biased tracers of the underlying matter density field. More recently however the gravitational deflection of light has also been shown to be usable in order to map the large scale distribution of dark matter, having the additional advantage of being insensitive to the problems related with the bias of luminous tracers. The tradeoff for this advantage is that cosmic shear can measure only a projected version of the matter power spectrum, that depends on the assumed redshift distribution of the background source galaxies.

In this paper we focused on this approach, namely we aimed at understanding what kind of constraints can be put on deviations from primordial Gaussianity using the weak lensing power spectrum. As an example, attention was devoted to planned wide field optical surveys, such as the ESA Cosmic Vision project EUCLID \citep{LA09.1}. The rest of this work is organized as follows. In Section \ref{sct:ng} we describe the non-Gaussian cosmologies considered in this work and how deviations from Gaussianity alter the mass function and the halo bias, both required for computing the non-linear power spectrum. In Section \ref{sct:modeling} we discuss in detail the way in which we modeled the fully non-linear matter power spectrum, with particular attention to the assumptions, advantages and drawbacks underlying the method. In Section \ref{sct:results} we describe our results concerning the weak lensing power spectrum, and in Section \ref{sct:discussion} we draw our conclusions.

For the relevant calculations we adopted as a reference cosmology the one resulting from the best fit WMAP-$5$ parameters together with type-Ia supernovae and the observed Baryon Acoustic Oscillation (BAO). The present values of the density parameters for matter, dark energy and baryons are $\Omega_{\mathrm{m},0} = 0.279$, $\Omega_{\Lambda,0} = 0.721$ and $\Omega_{\mathrm{b},0} = 0.046$, respectively. The Hubble constant reads $H_0 = h$100 km s$^{-1}$ Mpc$^{-1}$, with $h = 0.701$. The slope of the primordial power spectrum of density fluctuations is $n = 0.96$, while the normalization is set by the \emph{rms} of the density field on a comoving scale of $8 h^{-1}$ Mpc, $\sigma_8 = 0.817$. To construct the linear power spectrum we used the matter transfer function of \citet{BA86.1}, modified according to the shape factor of \citet{SU95.1}. The more sophisticated prescription of \citet{EI98.1} is almost coincident with the former, except for the presence of the BAO, that is not of interest here.

%%%%%%%%%%%%%%%%%%%%%%%%%%%%%%%%%%%%%%%%%%%%%%%%%%%
\section{Non-Gaussian cosmologies}\label{sct:ng}
%%%%%%%%%%%%%%%%%%%%%%%%%%%%%%%%%%%%%%%%%%%%%%%%%%%

Simple generalizations of the most standard model of inflation give rise to seed primordial density fluctuations that follow a non-Gaussian probability distribution. A particularly simple way to parametrize the deviation of this distribution from a Gaussian consists of writing the Bardeen's gauge invariant potential $\Phi$ as the sum of a linear Gaussian term and a quadratic correction \citep{SA90.1,GA94.1,VE00.1,KO01.1}, 

\begin{equation}\label{eqn:ng}
\Phi = \Phi_\mathrm{G} + f_\mathrm{NL}*\left( \Phi_\mathrm{G}^2 - \langle \Phi_\mathrm{G}^2 \rangle \right).
\end{equation}
In Eq. (\ref{eqn:ng}) the symbol $*$ denotes convolution between functions, and reduces to simple multiplication only in the particular case in which $f_\mathrm{NL}$ is a constant, while in general it is a function of the scale. Note that on scales smaller than the Hubble radius $\Phi$ equals minus the Newtonian peculiar gravitational potential.

In the following, we adopted the Large Scale Structure convention (as opposed to the CMB convention, see \citealt*{AF08.1,PI09.1,CA08.1}; \citealt{GR09.1}) for defining the fundamental parameter $f_\mathrm{NL}$. According to this, the primordial value of $\Phi$ has to be linearly extrapolated at $z = 0$, and as a consequence the constraints given on $f_\mathrm{NL}$ by the CMB have to be raised of $\sim 30\%$ to comply with this paper's convention (see also \citealt*{FE09.1} for a concise explanation).

If the distribution of the primordial density (and potential) perturbations is not Gaussian, then it cannot be fully described by the power spectrum $P_\Phi(\bf{k})$ only, but we also need higher-order moments such as the bispectrum $B_\mathrm{\Phi}({\bf k}_1,{\bf k}_2,{\bf k}_3)$. In particular, different models of inflation give rise to different shapes of the bispectrum. In the following we shall adopt two particularly popular bispectrum shapes. The first one is dubbed the \emph{local} shape. In this case, the bispectrum is maximized for configurations in which one of the three momenta is much smaller than the other two ("squeezed" configurations). The parameter $f_\mathrm{NL}$ is a dimensionless constant, and the bispectrum can be written as \citep{CR07.1,LO08.1}

\begin{equation}
B_\Phi({\bf k}_1,{\bf k}_2,{\bf k}_3) = 2f_\mathrm{NL} B^2 \left[ k_1^{n-4}k_2^{n-4} + k_1^{n-4}k_3^{n-4} + k_2^{n-4}k_3^{n-4} \right],
\end{equation}
where $k_i = \|{\bf k}_i\|$. The constant $B$ is the amplitude of the spectrum $P_\Phi(k)$, related to the amplitude $A$ of the power spectrum of density fluctuations by the relation $B = 9AH_0^4\Omega_{\mathrm{m},0}^2/4$. The second bispectrum shape is the \emph{equilateral} shape, where the bispectrum is maximized by configurations where the three arguments have approximately the same magnitude. In the latter case, the primordial bispectrum takes the cumbersome form

\begin{eqnarray}
B_\Phi({\bf k}_1,{\bf k}_2,{\bf k}_3) &=&  6f_\mathrm{NL} B^2  \left[ k_1^{(n-4)/3}k_2^{2(n-4)/3}k_3^{n-4} \right. + 
\nonumber\\
&+& k_3^{(n-4)/3}k_1^{2(n-4)/3}k_2^{n-4} + k_2^{(n-4)/3}k_3^{2(n-4)/3}k_1^{n-4} + 
\nonumber\\
&+& k_1^{(n-4)/3}k_3^{2(n-4)/3}k_2^{n-4} + k_2^{(n-4)/3}k_1^{2(n-4)/3}k_3^{n-4} + 
\nonumber\\
&+& k_3^{(n-4)/3}k_2^{2(n-4)/3}k_1^{n-4} -k_1^{n-4}k_2^{n-4} - k_1^{n-4}k_3^{n-4} -
\nonumber\\
&-& \left. k_2^{n-4}k_3^{n-4} - 2k_1^{2(n-4)/3}k_2^{2(n-4)/3}k_3^{2(n-4)/3}\right]. 
\end{eqnarray}
Most importantly, in inflationary models that predict an equilateral primordial bispectrum, the parameter $f_\mathrm{NL}$ is in general dependent on the scales. We adopt here the functional form suggested by \citet{LO08.1}, according to which

\begin{equation}\label{eqn:kcmb}
f_\mathrm{NL}({\bf k}_1,{\bf k}_2,{\bf k}_3) = f_{\mathrm{NL},0} \left( \frac{k_1+k_2+k_3}{k_\mathrm{CMB}} \right)^{-2\kappa}.
\end{equation}
The functional form of Eq. (\ref{eqn:kcmb}) is chosen in order to avoid violating the WMAP constraints. Specifically, $f_{\mathrm{NL},0}$ represents the non-linear parameter evaluated at the scale $k_\mathrm{CMB} = 0.086 h$ Mpc$^{-1}$ roughly corresponding to the largest multipole used by \cite{KO09.1} to estimate non-Gaussianity in the WMAP data, $l = 700$. The constant free parameter $\kappa$ is assumed to be $|\kappa| \ll 1$ between CMB and cluster scales \citep{LO08.1,CR09.1}, in order to enhance non-Gaussianity on scales smaller than CMB. In previous work (\citealt*{FE09.1}), we adopted the values $\kappa = 0, -0.1, -0.2$. Here, for simplicity we limit ourselves to the case $\kappa = -0.2$, that is expected to give the largest effect. Also, for ease of notation, in the equilateral case we shall henceforth write $f_\mathrm{NL}$ meaning $f_{\mathrm{NL},0}$, since no ambiguity will arise.

At least two of the ingredients that make up the non-linear matter power spectrum (Section \ref{sct:modeling}) are modified in cosmologies with non-Gaussian initial conditions: the halo mass function and the linear bias. The mass function $n(M,z)$ is the number of structures within the unit mass around $M$ that at redshift $z$ is contained in the unit comoving volume. An often used prescription for the mass function in Gaussian cosmologies is the one of \cite{PR74.1}, that however tends to overpredict halo abundance at low masses with respect to the results of numerical $n$-body simulations. Other prescriptions exist, whose parameters are fitted against $n$-body simulations \citep{JE01.1,WA06.1,TI08.1} or determined based on more realistic models for the collapse of density perturbations \citep*{SH01.1,SH02.1}. We used the latter prescription in the remainder of this work. The \citet{SH02.1} mass function in non-Gaussian models can then be written as

\begin{equation}
n(M,z) = n^{\mathrm{(G)}}(M,z) \frac{n_\mathrm{PS}(M,z)}{n_\mathrm{PS}^\mathrm{(G)}(M,z)},
\end{equation}
where in our case $n^{\mathrm{(G)}}(M,z)$ is the mass function in the Gaussian cosmology computed according to the \cite{SH02.1} prescription and $n_\mathrm{PS}(M,z)$ and $n_\mathrm{PS}^\mathrm{(G)}(M,z)$ represent the \cite{PR74.1} mass functions in the non-Gaussian and reference Gaussian models respectively. Following \cite{LO08.1}, we write the mass function $n_\mathrm{PS}(M,z)$ as

\begin{eqnarray}\label{eqn:mfps}
n_\mathrm{PS}(M,z) &=& - \sqrt{\frac{2}{\pi}} \frac{\rho_\mathrm{m}}{M} \exp\left[ -\frac{\delta_\mathrm{c}^2(z)}{2\sigma_M^2} \right] \left[ \frac{d\ln \sigma_M}{dM} \left( \frac{\delta_\mathrm{c}(z)}{\sigma_M} + \right.\right.
\nonumber\\
&+& \left. \left. \frac{S_3\sigma_M}{6} \left( \frac{\delta_\mathrm{c}^4(z)}{\sigma^4_M} -2\frac{\delta^2_\mathrm{c}(z)}{\sigma^2_M} -1\right) \right) + \right.
\nonumber\\
&+& \left. \frac{1}{6} \frac{dS_3}{dM}\sigma_M \left( \frac{\delta^2_\mathrm{c}(z)}{\sigma^2_M} -1\right) \right].
\end{eqnarray}

Eq. (\ref{eqn:mfps}) has been obtained by Edgeworth expanding the probability distribution for smoothed density fluctuations and $\delta_\mathrm{c}(z) \equiv \Delta_\mathrm{c}/D_+(z)$, where the quantity $\Delta_\mathrm{c}$ is the linear density threshold for spherical collapse, that is constant in an Einstein-de Sitter model and only mildly dependent on redshift in models with a cosmological constant. We included this redshift dependence in our calculations but do not indicate it explicitely, since it is practically irrelevant. The function $D_+(z)$ is the linear growth factor for density fluctuations, while $\sigma_M$ is the \emph{rms} of density perturbations smoothed on a scale corresponding to mass $M$.
The function $S_3(M) \equiv f_{\mathrm{NL},0} \mu_3(M)/\sigma_M^4$ is the normalized skewness. Note that Eq. (\ref{eqn:mfps}) reduces to the standard $n_\mathrm{PS}^\mathrm{(G)}(M,z)$ in the case in which $S_3(M)$ vanishes identically. The third-order moment $\mu_3(M)$ can be written as

\begin{equation}
\mu_3(M) = \int_{\mathbb{R}^9} \mathcal{M}_R(k_1) \mathcal{M}_R(k_2) \mathcal{M}_R(k_3) B_\Phi({\bf k}_1,{\bf k}_2,{\bf k}_3) \frac{d{\bf k}_1d{\bf k}_2d{\bf k}_3}{(2\pi)^9}.
\end{equation}
The function $\mathcal{M}_R(k)$ relates the Fourier transform of density  fluctuations smoothed on some scale $R$ to the relative peculiar potential, and it is defined as

\begin{equation}
\mathcal{M}_R(k) \equiv \frac{2}{3} \frac{T(k)k^2}{H_0^2 \Omega_{\mathrm{m},0}} W_R(k),
\end{equation}
where $T(k)$ is the matter transfer function and $W_R(k)$ is the Fourier transform of the top-hat window function. For an alternative derivation of $n_\mathrm{PS}(M,z)$, see \citet*{MA00.2}.

The linear bias describes how well dark matter halos trace the underlying large scale matter distribution, and is needed in order to account for the correlation between different halos. For it we adopted the \citet*{SH01.1} modification of the original \citet{MO96.1} formula, obtained with \cite{PR74.1}-like considerations in Gaussian cosmologies, that reads 

\begin{eqnarray}\label{eqn:bias}
b^\mathrm{(G)}(M,z) &=& 1 + a\frac{\Delta_\mathrm{c}}{D_+^2(z)\sigma^2_M} - \frac{1}{\Delta_\mathrm{c}} +
\nonumber\\
&+& \frac{2p}{\Delta_\mathrm{c}} \left[ \frac{[D_+(z)\sigma_M]^{2p}}{[D_+(z)\sigma_M]^{2p} + [\sqrt{a} \Delta_\mathrm{c}]^{2p}} \right].
\end{eqnarray}
The original \citet{MO96.1} formula is obtained by setting $a = 1$ and $p = 0$ in Eq. (\ref{eqn:bias}), while the \citet*{SH01.1} revision is obtained with the values $a = 0.75$ and $p = 0.3$.

In non-Gaussian models, the bias acquires an extra scale dependence that can be written as \citep{MA08.1}

\begin{equation}
b(M,z,k) = b^\mathrm{(G)}(M,z) + \Delta b(M,z,k),
\end{equation}
where 

\begin{equation}
\Delta b(M,z,k) = \left[ b^\mathrm{(G)}(M,z)-1 \right] \delta_\mathrm{c}(z) \Gamma_R(k).
\end{equation}
The term $\Gamma_R(k)$ encapsulates all the dependence on the scale, and can be written as

\begin{eqnarray}
\Gamma_R(k) &=& \frac{1}{8\pi^2\sigma_M^2\mathcal{M}_R(k)} \int_0^{+\infty} \zeta^2\mathcal{M}_R(\zeta) \times
\nonumber\\
&\times& \left[ \int_{-1}^1 \mathcal{M}_R\left(\sqrt{\alpha}\right) \frac{B_\Phi\left( \zeta,\sqrt{\alpha},k \right)}{P_\Phi(k)} d\mu \right] d\zeta,
\end{eqnarray}
where $\alpha \equiv k^2 + \zeta^2 + 2k\zeta\mu$ and $R$ is the top-hat radius corresponding to the mass $M$. In the particular case of a primordial bispectrum of local shape, the previous equation simplifies to

\begin{eqnarray}
\Gamma_R(k) &=& \frac{2f_\mathrm{NL}}{8\pi^2\sigma_M^2\mathcal{M}_R(k)} \int_0^{+\infty} \zeta^2\mathcal{M}_R(\zeta) P_\Phi(\zeta)\times
\nonumber\\
&\times& \left[ \int_{-1}^1 \mathcal{M}_R\left(\sqrt{\alpha}\right) \frac{P_\Phi\left(\sqrt{\alpha}\right)}{P_\Phi(k)} d\mu \right] d\zeta.
\end{eqnarray}

In the next section we show how the mass function and halo bias enter the non-linear matter power spectrum, and what is the subsequent effect of primordial non-Gaussianity.

%%%%%%%%%%%%%%%%%%%%%%%%%%%%%%%%%%%%%%%%%%%%%%%%%%%
\section{Modeling the non-linear power spectrum}\label{sct:modeling}
%%%%%%%%%%%%%%%%%%%%%%%%%%%%%%%%%%%%%%%%%%%%%%%%%%%

We computed the fully non-linear matter power spectrum by using the halo model developed by \citet{MA00.3} and \citet{SE00.1}. In this model, the power spectrum is set by the sum of two terms. The first one, dominating on large scales, is given by dark-matter particle pairs residing in different halos, hence it depends on the correlations of individual halos. The second term, dominating on the smallest scales, takes into account particle pairs that are included in the same halo, hence it is extremely sensitive to the inner structure of halos themselves. This kind of decomposition of the matter power spectrum probably has a deeper rooting than simple power spectrum modeling, since it also arises in renormalized perturbation theory \citep{CR06.2,CR06.1}. 

The main ingredients entering in this model are the mass function, the halo bias and the halo internal structure. The first two were discussed in Section \ref{sct:ng}, including their modifications due to non-Gaussian initial conditions. The only additional point that we make here is that one of the hypotheses underlying the halo model for the power spectrum is that all the matter in the Universe is included inside halos of some mass. This imposes the constraint

\begin{equation}\label{eqn:c1}
\int_0^{+\infty} n(M,z) \frac{M}{\rho_\mathrm{m}} dM = 1,
\end{equation}
where $\rho_\mathrm{m}$ is the average comoving matter density. As can be easily verified, this constraint is fulfilled in Gaussian cosmologies by the \cite{PR74.1} mass function, and also the \cite{SH02.1} mass function is normalized such to satisfy Eq. (\ref{eqn:c1}). Numerically however it is not possible to push the lower integration bound down to zero, rather the integration will be stopped to some $M_\mathrm{inf} > 0$. As noted by \cite{RE02.2},  decreasing $M_\mathrm{inf}$ makes the integral on the left-hand side of Eq. (\ref{eqn:c1}) approach unity, however this happens very slowly in CDM models, so that effectively, even if $M_\mathrm{inf}$ is very small, the integral will still be significantly different from unity. We solved this following \cite{RE02.2}, namely by enforcing the constraint in Eq. (\ref{eqn:c1}) by adding a constant to the mass function in the smallest mass bin that is considered when performing the integration numerically. In the present work, we assumed $M_\mathrm{inf} = 10^6 M_\odot h^{-1}$.

%%%%%%%%%%%%%%%%%%%%%%%%%%%%%%%%%%%%%%%%%%%%%%%%%%%
\subsection{Halo density profile}
%%%%%%%%%%%%%%%%%%%%%%%%%%%%%%%%%%%%%%%%%%%%%%%%%%%

We assumed that dark matter halos in both Gaussian and non-Gaussian cosmological models are on average well described by a generalized \citet*{NA96.1} (NFW henceforth, see also \citealt*{NA95.1,NA97.1}) density profile, written as

\begin{equation}
\rho(r) = \frac{\rho_\mathrm{s}}{(r/r_\mathrm{s})^\alpha(1+r/r_\mathrm{s})^{3-\alpha}},
\end{equation}
which reduces to the standard NFW model for $\alpha = 1$ \citep{AM04.1}. For full generality we developed the following calculations for any value of $\alpha$, however when evaluating the effect of non-Gaussianity on the power spectrum we specialized mostly to the case $\alpha = 1$, with only a minor discussion on the role of the inner slope. The total mass of the halo that is included inside some radius $r$ can be computed as \citep{TA03.2}

\begin{eqnarray}\label{eqn:mass}
M(r) &=& 4\pi \int_0^r \rho(x) x^2 dx = 
\nonumber\\
&=&\frac{4\pi \rho_\mathrm{s}r_\mathrm{s}^3}{3-\alpha} \left._2F_1 \right. \left( 3-\alpha,3-\alpha;4-\alpha;-\frac{r}{r_\mathrm{s}} \right) \left( \frac{r}{r_\mathrm{s}} \right)^{3-\alpha} \equiv
\nonumber\\
&\equiv& 4\pi \rho_\mathrm{s}r_\mathrm{s}^3G_\alpha(r/r_\mathrm{s}).
\end{eqnarray}
In Eq. (\ref{eqn:mass}), the function $\left._2F_1 \right.$ is the Gauss hypergeometric function, and in the particular case $\alpha = 1$, the function $G_\alpha$ reduces to the well known form

\begin{equation}
G_1(x) = \ln(1+x) - \frac{x}{1+x}.
\end{equation}

The two parameters $r_\mathrm{s}$ and $\rho_\mathrm{s}$ completely specify the density profile and can be expressed in terms of the virial mass $M$ and concentration $c$. In the remainder of this work we defined the virial mass as the mass contained in the sphere whose mean density equals $\Delta_\mathrm{v} = 200$ times the \emph{average} density of the Universe, that is 

\begin{equation}\label{eqn:massv}
M = \frac{4}{3}\pi R_\mathrm{v}^3 \Delta_\mathrm{v} \rho_\mathrm{m}.
\end{equation}
The virial radius $R_\mathrm{v}$ is the radius of this sphere, and the concentration is defined as the ratio between the virial radius and the scale radius of the profile, $c \equiv R_\mathrm{v}/r_\mathrm{s}$. It is important to note that different authors use different definitions for the viral radius. In some cases the overdensity is not referred to the mean matter density, rather to the critical density $\rho_\mathrm{c}(z) = 3H^2(z)/8\pi G$. Yet others use different values of the overdensity, usually the value computed for the collapse of a spherical overdensity, that in an Einstein-de Sitter universe is constant and equals $\Delta_\mathrm{v} \simeq 178$. It is expected that different choices assign different values of the concentration to the same virial mass, hence effectively modifying the power spectrum on small scales. We adopted the overdensity with respect to the average density for consistency with the \cite{SH02.1} mass function, that was calibrated against numerical simulations where dark matter lumps were detected via spherical overdensity methods \citep{TO98.1}. They actually used $\Delta_\mathrm{v} = 178$, but we checked that this does not make a significant difference for the resultant power spectrum.

Given the previous considerations, the comoving scale density can then be written as

\begin{equation}
\rho_\mathrm{s} = \frac{\Delta_\mathrm{v}}{3} \rho_\mathrm{m} \frac{c^3}{G_\alpha(c)},
\end{equation}
and the comoving scale radius as

\begin{equation}
r_\mathrm{s} = \left( \frac{3M}{4\pi c^3 \Delta_\mathrm{v}\rho_\mathrm{m}} \right)^{1/3}.
\end{equation}
Please note that the scale radius is independent on the slope $\alpha$, while the scale density is not. 

The concentration of a dark matter halo is actually linked to the virial mass according to the hierarchical paradigm for structure formation, since it is expected that small structures collapse earlier and hence are more compact at a given redshift. We discuss the exact nature of this relationship further below. Thus, once the inner slope of the profile $\alpha$ is specified, the dark matter distribution depends effectively only on mass and redshift. Hence, for a dark matter halo of mass $M$ at redshift $z$ from this moment on we write the density profile as $\rho(r,M,z)$. We indicate with $\hat{\rho}(k,M,z)$ the Fourier transform of $\rho(r,M,z)$ with respect to the radius, which can be written as

\begin{equation}\label{eqn:ft}
\hat{\rho}(k,M,z) = 4\pi \int_0^{R_\mathrm{v}} \rho(r,M,z) \frac{\sin(kr)}{kr}r^2dr.
\end{equation}
This definition conveniently implies $\hat{\rho}(0,M,z) = M$, although it neglects matter outside the virial radius (see the discussion below). When $\alpha = 1$ the Fourier transform of the density profile can be expressed analitically, according to 

\begin{figure*}
	\includegraphics[width=0.45\hsize]{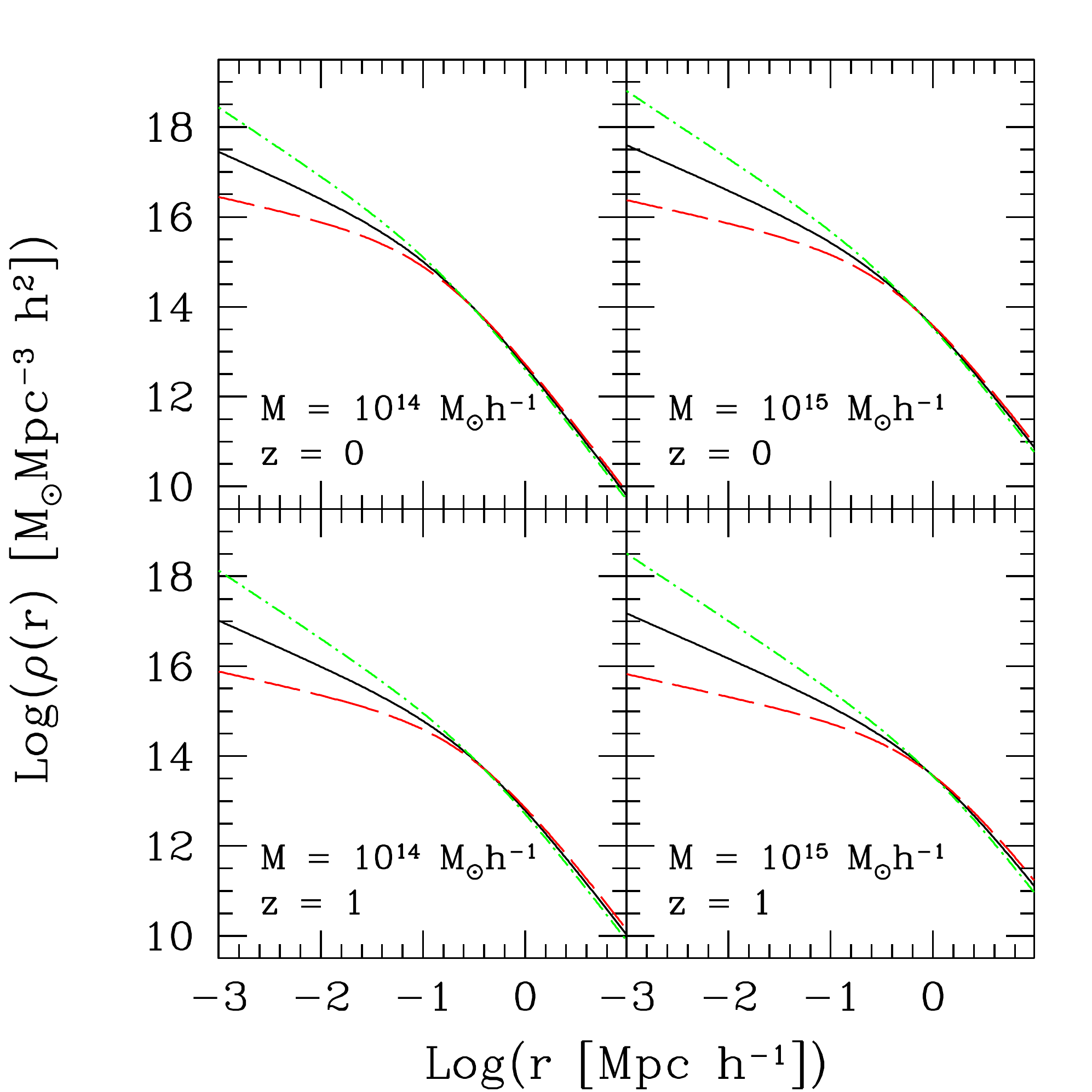}
	\includegraphics[width=0.45\hsize]{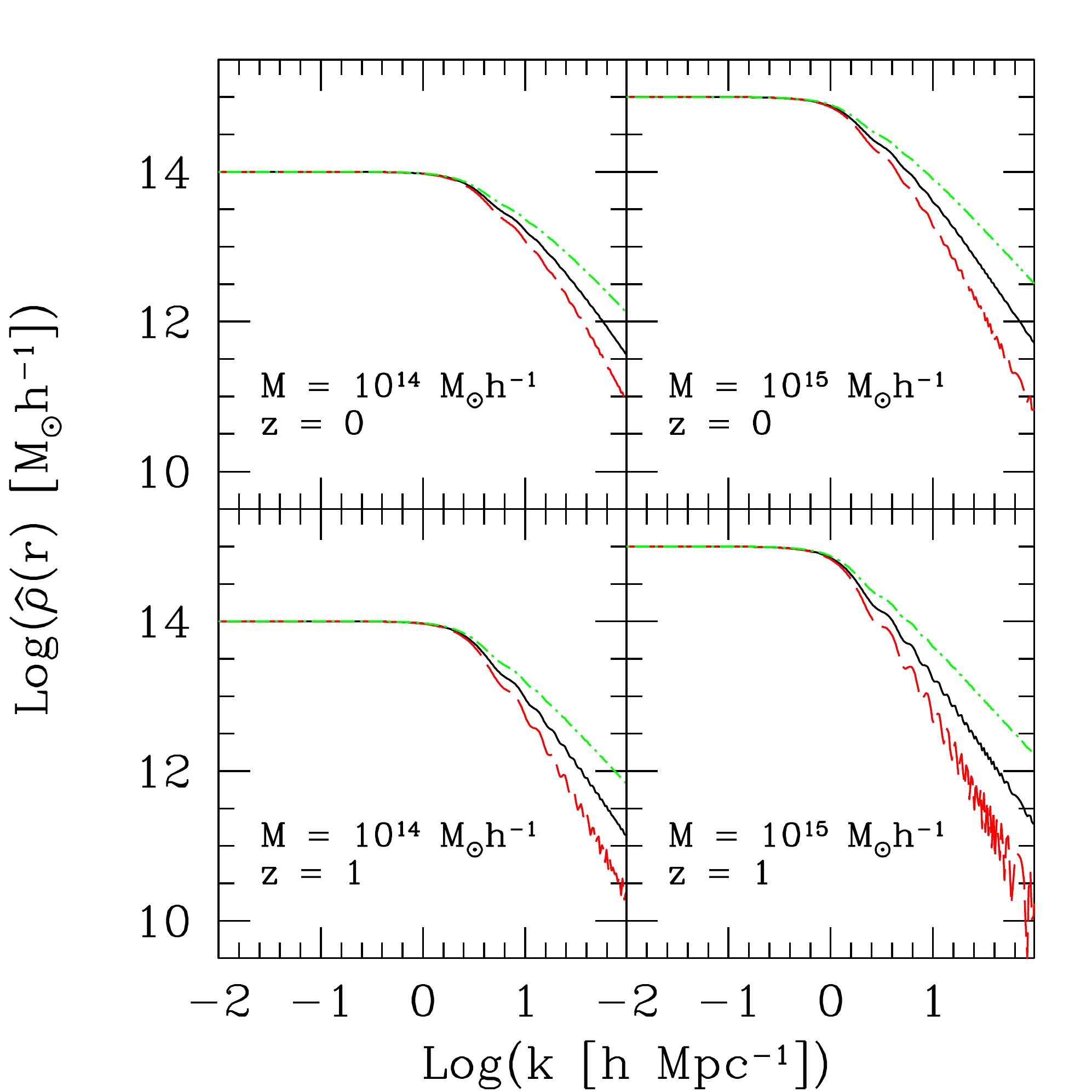}\hfill
	\caption{\emph{Left panel}. The density profiles of dark matter halos with different masses and redshifts, as labelled in the various panels. Different line styles and colors refer to different slopes of the inner density profiles, namely $\alpha = 0.5$ (red long-dashed), $\alpha = 1$ (black solid line) and $\alpha = 1.5$ (green dot-dashed line). \emph{Right panel}. The Fourier transform of the same density profiles shown in the left panel. The concentration was related to the virial mass via the prescription of \citet*{EK01.1}. Note that at small wavenumbers the Fourier transform of the density profile correctly equals the virial mass.}
\label{fig:profile}
\end{figure*}

\begin{eqnarray}
\hat{\rho}(k,M,z) &=& 4\pi \rho_\mathrm{s}r_\mathrm{s}^3 \left[\frac{}{}\sin(kr_\mathrm{s}) \left(\mathrm{Si}[(1+c)kr_\mathrm{s}] - \mathrm{Si}(kr_\mathrm{s}) \right) \right. + 
\nonumber\\
&+&\cos(kr_\mathrm{s}) \left(\mathrm{Ci}[(1+c)kr_\mathrm{s}] - \mathrm{Ci}(kr_\mathrm{s}) \right) -
\nonumber\\
&-&\left.\frac{\sin(kr_\mathrm{s}c)}{(1+c)kr_\mathrm{s}} \right]
\end{eqnarray}
(\citealt{SC01.1}; \citealt*{RU08.2}), where $\mathrm{Si}(x)$ and $\mathrm{Ci}(x)$ are the sine and cosine integrals respectively. When $\alpha$ is different from unity, we have instead to resort to numerical integration. Please note that, for $x \rightarrow 0$, $\mathrm{Si}(x) \simeq 0$ while $\mathrm{Ci}(x) \simeq \ln(x)$, thus we have that $\hat{\rho}(0,M,z) = 4\pi \rho_\mathrm{s}r_\mathrm{s}^3 G_1(c)$, that, according to Eq. (\ref{eqn:mass}), correctly equals the virial mass of the halo.

In Figure \ref{fig:profile} we show the density profiles of dark matter halos with different masses and redshifts and their Fourier transforms. We plot results for three different values of the inner slope $\alpha$, and we adopted the \citet*{EK01.1} prescription in order to relate the mass to the concentration of dark matter halos. We discuss this choice further below. Due to the sine function present in the integral in Eq. (\ref{eqn:ft}), the Fourier transform of the density profile presents wiggles at small scales, whose strength increases with decreasing $\alpha$. This can be naively understood since profiles with smaller $\alpha$ are flatter, and hence it is expected that their Fourier transforms have more fluctuations at large wavenumbers.

%%%%%%%%%%%%%%%%%%%%%%%%%%%%%%%%%%%%%%%%%%%%%%%%%%%
\subsection{Power spectrum}
%%%%%%%%%%%%%%%%%%%%%%%%%%%%%%%%%%%%%%%%%%%%%%%%%%%

Let now $P_\mathrm{L}(k,z)$ be the linear power spectrum of density fluctuations extrapolated at redshift $z$. We assumed it to be the same both in the reference Gaussian model and in non-Gaussian cosmologies. According to the halo model, the fully non-linear spectrum $P(k,z)$ can be written as the sum of two terms that read as follows.

\begin{equation}\label{eqn:p1}
P_1(k,z) = \int_0^{+\infty} n(M,z) \left[\frac{\hat{\rho}(M,z,k)}{\rho_\mathrm{m}}\right]^2 dM
\end{equation}
and
\begin{equation}\label{eqn:p2}
P_2(k,z) = \left[ \int_0^{+\infty} n(M,z)b(M,z,k)\frac{\hat{\rho}(M,z,k)}{\rho_\mathrm{m}}dM \right]^2 P_\mathrm{L}(k,z).
\end{equation}
In the two previous equations, $n(M,z)$ is the standard mass function while $b(M,z,k)$ is the linear bias, both introduced in Section \ref{sct:ng} above.

The full non-linear power spectrum on large scales is dominated by the second term , $P(k,z) \simeq P_2(k,z)$. As noted by \cite{SE00.1}, for self-consistency it is necessary that this term approaches the linear power spectrum in the limit $k \ll 1 h$ Mpc$^{-1}$, which imposes the nontrivial constraint

\begin{equation}\label{eqn:c2}
\int_0^{+\infty} n(M,z)b(M,z,k)\frac{M}{\rho_\mathrm{m}}dM = 1
\end{equation}
in that limit. Analogously to the previous mass function constraint in Eq. (\ref{eqn:c1}), we practically enforced the constraint in Eq. (\ref{eqn:c2}) by adding a constant to the bias in the smallest mass bin adopted in the numerical integration. There is one point worth of discussion about Eq. (\ref{eqn:c2}). While in the Gaussian model this constraint can be computed only once, given the redshift, in a non-Gaussian cosmology this computation has to be performed at each scale at which we are interested in computing the power spectrum. Strictly speaking, the condition needs to be enforced only in the limit $k \ll 1 h$ Mpc$^{-1}$, therefore we would have the freedom to relax it for $k \gtrsim 1 h$ Mpc$^{-1}$. However, it is not clear at which scale the transition between corrected and uncorrected bias should happen, neither how fast this transition should be. Therefore, we chose not to use this freedom, and to enforce the condition in Eq. (\ref{eqn:c2}) for all $k$ in non-Gaussian models.

In the remainder of this paper, unless explicitly noted, instead of the power spectrum we will refer to the dimensionless power $\Delta^2(k,z)$, defined as

\begin{equation}
\Delta^2(k,z) \equiv \frac{4\pi k^3 P(k,z)}{(2\pi)^3}.
\end{equation}

The last thing that remains to be defined in order to fully specify the halo model is the relation between the virial mass and the concentration of dark matter halos. This is of fundamental importance at very small scales, where the power spectrum is expected to be dominated by the correlations of dark matter particle pairs that are inside the same halo. In order to do this, there exist prescriptions based on the study of samples of dark matter halos extracted from high-resolution Gaussian cosmological simulations (\citealt*{NA96.1}; \citealt{BU01.1}; \citealt*{EK01.1}; \citealt{DO04.1,GA08.1}). Many authors however \citep*{CO00.2,RE02.2}, prefer to adopt a concentration-mass relation for which the halo model matter power spectrum is a good fit to the spectrum measured in $n$-body simulations of Gaussian cosmologies. The latter is often assumed to be well represented by the prescriptions of \cite{PE94.1} and of \citet{SM03.1}, both of which are implemented by the publicly available \texttt{halofit} code. 

\begin{figure}
	\includegraphics[width=\hsize]{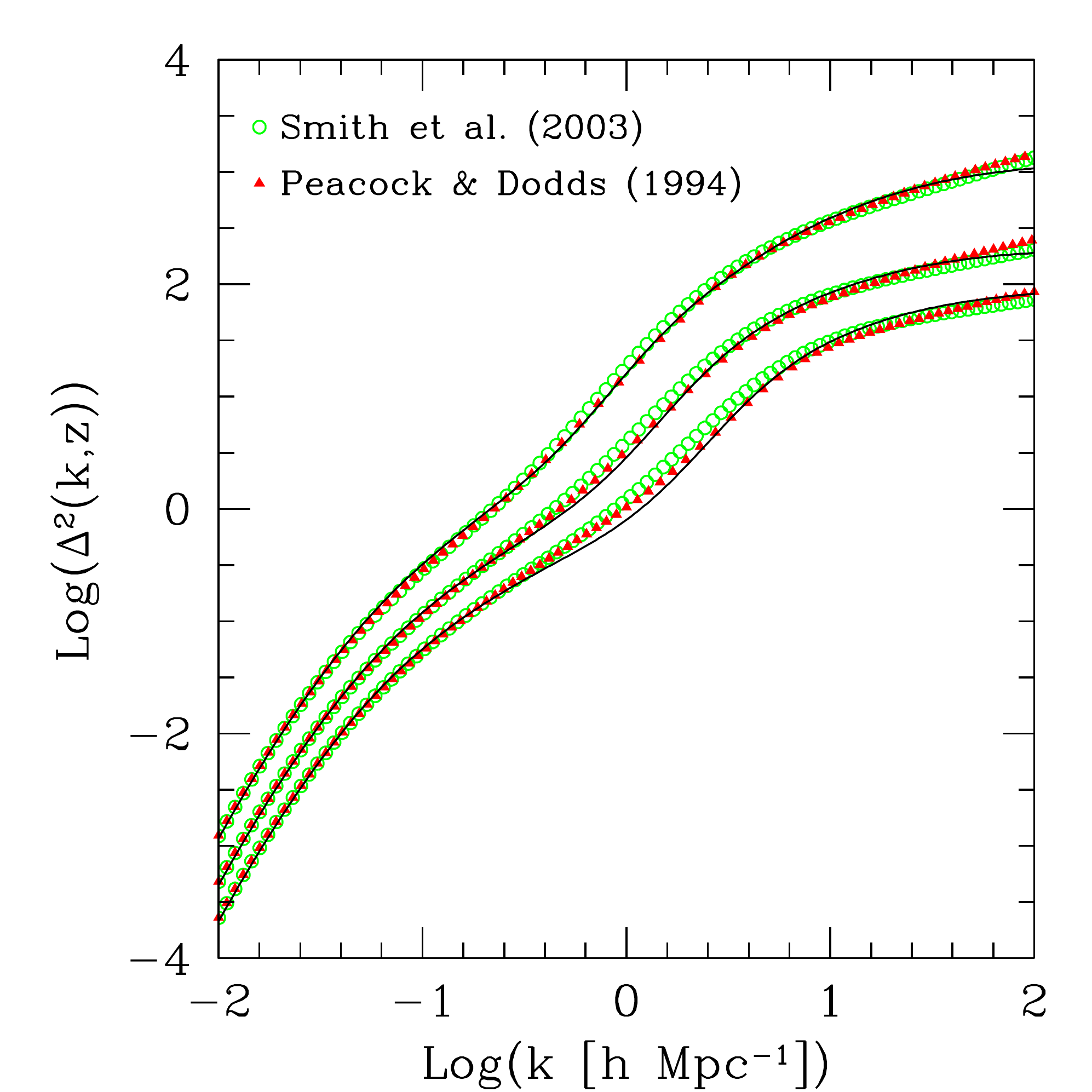}\hfill
	\caption{The dimensionless power computed according to the halo model in the reference Gaussian cosmology for three different redshifts, $z = 0$, $z = 1$ and $z = 2$, from top to bottom respectively. Results are compared with the recipes of \citet{PE94.1} and \citet{SM03.1} computed with \texttt{halofit}, as labeled. The concentration-mass relation is the one prescribed by \citet*{EK01.1}, with concentrations multiplied by the additional factor $(1+z)^{-1/2}$.}
	\label{fig:spectraZ}
\end{figure}

In a perfectly consistent picture of structure formation, the two described approaches should give equivalent results. In practice however, this is not the case. As noted by \cite{SE03.2}, in order for the first approach to reproduce numerically simulated spectra, the amplitude of the concentration-mass relation adopted should drop with redshift more steeply than predicted by both the \citet{BU01.1} and \citet*{EK01.1} prescriptions. \cite{SE03.2} argue that the halo model with a concentration-mass relation derived directly by simulated dark matter halos should be adopted as physically motivated, while the fits of \cite{PE94.1} and \cite{SM03.1} to numerically simulated power spectra should not be trusted beyond the range where they have been tested, that is $k \lesssim 40 h$ Mpc$^{-1}$ at $z = 0$ and even $k \lesssim 10 h$ Mpc$^{-1}$ at high redshift. We obtained results that are somewhat consistent with this interpretation. In fact, we found that by using the halo model with the concentration-mass relation given by \citet*{EK01.1} produces a $z = 0$ power spectrum that is in broad agreement with the \texttt{halofit} result. Moving at high redshift we find instead that the halo model power spectrum is higher than the \texttt{halofit} results for $k \gtrsim 10 - 20 h$ Mpc$^{-1}$, with the difference growing with redshift. If we require the concentration to drop with redshift more steeply than with the \citet*{EK01.1} recipe, as suggested by \cite{SE03.2}, we would reduce the power at small scales and high redshift, hence reducing the discrepancy. As a matter of fact, multiplying the \citet*{EK01.1} concentrations by the extra factor $(1+z)^{-1/2}$ we obtained a fair agreement, as shown in Figure \ref{fig:spectraZ}.

We also tried to use the concentration-mass relation that \cite{CO00.2} found to give a good fit to numerically simulated power spectra (see also \citealt{RE02.2}). We found that the resulting power at small scales is higher than the \texttt{halofit} results. This is likely a consequence of the fact that the fit of \cite{CO00.2} (as the authors themselves note) is valid only for the specific cosmological model they used, that in particular has a higher normalization $\sigma_8$ than our. This implies larger halo concentrations for fixed mass and redshift, and hence more power at large wavenumbers. Approaches similar in spirit to those of \cite{CO00.2} and \cite{RE02.2} have recently been followed by \citet*{BE09.1} in order to find a suitable version of the mass-concentration relation of dark matter halos.

A distinct possibility is that the fits to numerically simulated power spectra are indeed correct even beyond their range of applicability, but the halo profile to be inserted in the halo model is not the true profile of dark matter halos, since other effects such as halo substructure and triaxiality can affect the power spectrum on small scales \citep{CO00.2}. However, \cite{SE00.1} and \cite{SE03.2} argue that at least the effect of substructures is not significant.

\begin{figure}
	\includegraphics[width=\hsize]{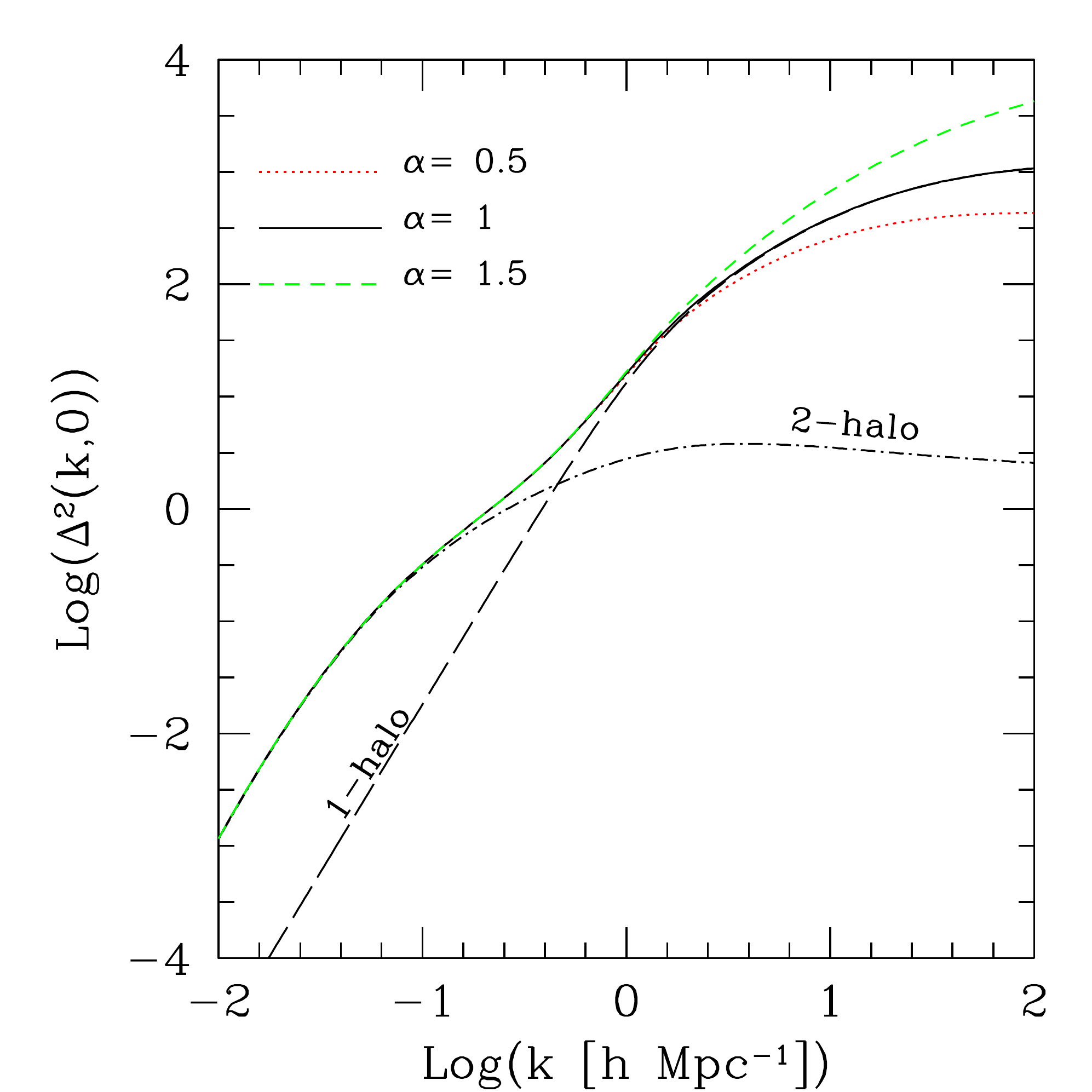}\hfill
	\caption{The fully non-linear dark matter dimensionless power in the reference Gaussian cosmology at $z = 0$ for three different values of the inner slope of dark matter halos, as labelled in the plot. Also, the long-dashed and dot-dashed curves represent the 1-halo and 2-halo contributions to the non-linear power spectrum for $\alpha = 1$. In all cases, the mass-concentration relation from \citet*{EK01.1} has been adopted, with the redshift correction needed to match the \texttt{halofit} results.}
	\label{fig:spectra}
\end{figure}

\begin{figure*}
	\includegraphics[width=0.45\hsize]{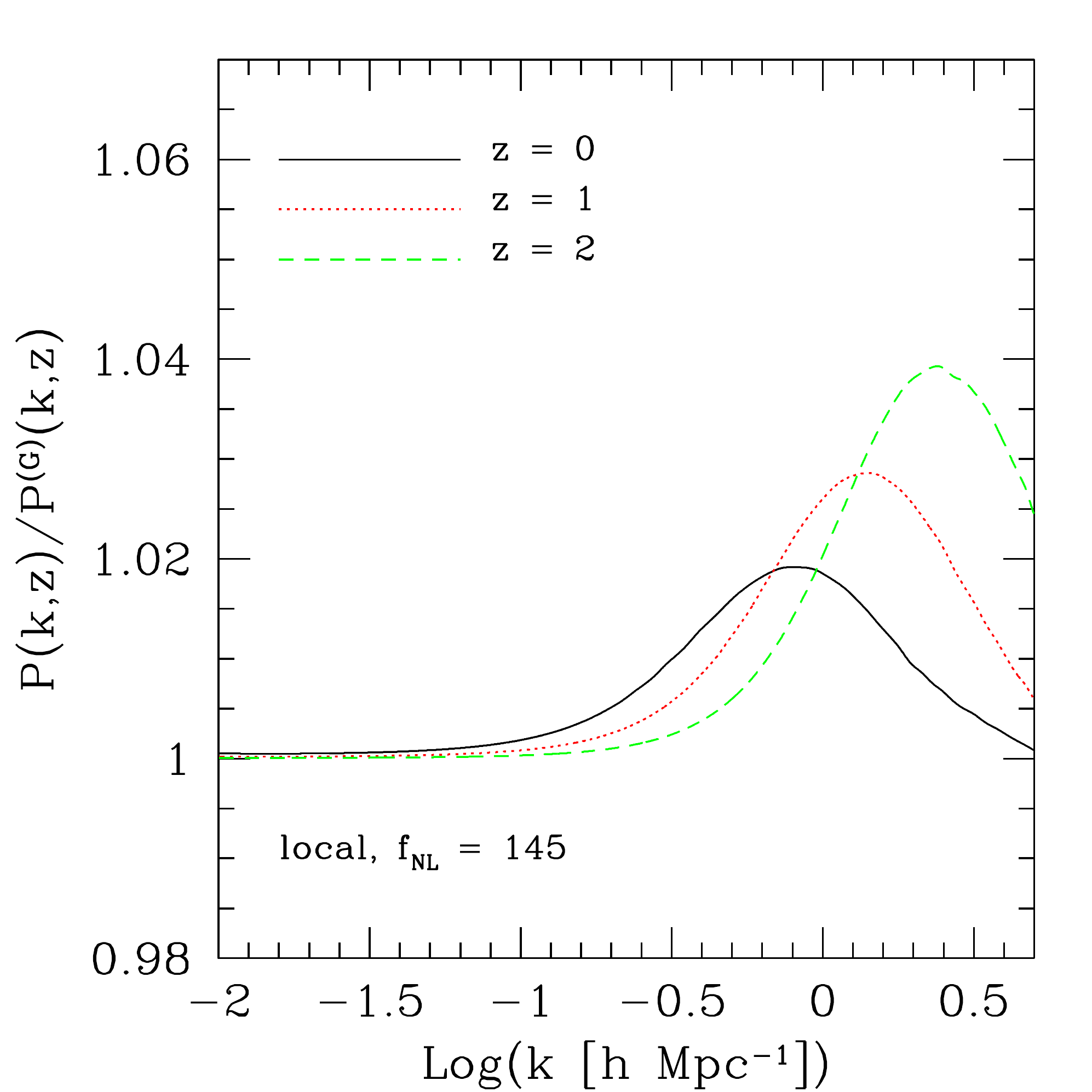}
	\includegraphics[width=0.45\hsize]{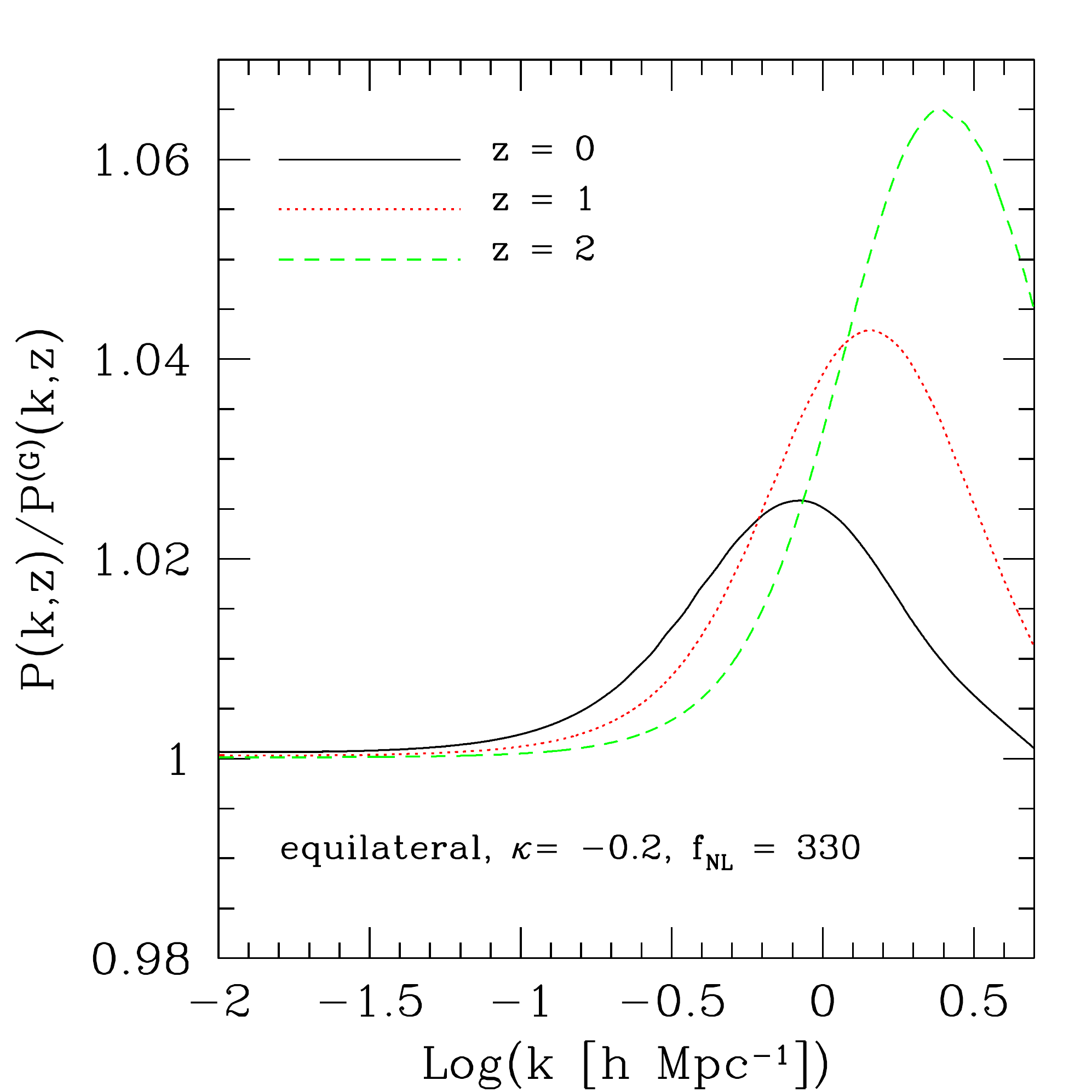}\hfill
	\caption{The ratio of the matter power spectrum computed in non-Gaussian cosmologies to the Gaussian one, for three different redshifts, as labeled. The left panel shows results for the local model with $f_\mathrm{NL} = 145$, while the right panel refer to a model with equilateral shape of the bispectrum, exponent of the scale dependence of the non-Gaussian amplitude $\kappa = -0.2$ (see Eq. \ref{eqn:kcmb}) and $f_\mathrm{NL} = 330$. Such numbers are the largest positive allowed by the WMAP data analysis of \citet{KO09.1}.}
	\label{fig:ratio}
\end{figure*}

Finally, as a last point we note that some degree of disagreement is also present in the intermediate regime in Figure \ref{fig:spectraZ}. This was also found and discussed by \cite{SE03.2}, who state that the halo model is not expected to be perfect at intermediate scales because of several factors, for instance the fact that we ignore the matter outside the virial radius in the Fourier transform in Eq. (\ref{eqn:ft}) and the contribution from the single halo term of the power spectrum is certainly overestimated for $k \ll 1/R_\mathrm{v}$. It is quite clear that with the precision of upcoming experiments for the measurement of the non-linear power spectrum, it is going to be necessary to find a unique prescription for the power spectrum, valid on all scales and in agreement with the mean structure of individual halos. However, for the time being we are interested in relative deviations with respect to the fiducial Gaussian $\Lambda$CDM cosmology, hence we stick to our choice.

In Figure \ref{fig:spectra} we show the fully non-linear power spectrum of the dark matter in the Gaussian cosmological model for three different values of the inner slope of dark matter halos. The concentrations are computed again according to the prescription of \citet*{EK01.1} with the correction factor $(1+z)^{-1/2}$, and for the case $\alpha = 1$ the separate contributions from the $1$-halo and $2$-halo terms are also shown. As it is expected, the inner slope mainly affects the power spectrum at large wavenumbers, with the dimensionless power increasing with increasing $\alpha$. The effect of the inner slope of halo density profiles on the power spectrum for the rather extreme values shown in Figure \ref{fig:spectra} can be quite large at small scales, and in fact we show below that for large enough wavenumbers it does overcome the effect of primordial non-Gaussianity captured by the halo model.

For realistic models of non-Gaussianity, it turns out that the shift of the matter power spectrum due to the corrections of the bias and of the mass function is very small, such that it is almost not visible on the scale of e.g., Figure \ref{fig:spectra} (see for instance \citealt{GR08.2}). Hence, in Figure \ref{fig:ratio}, we plot the ratio of the matter power spectrum for non-Gaussian cosmologies to the Gaussian one, with all quantities computed adopting $\alpha=1$. Shown are the results for primordial bispectra of the local and equilateral shapes (with $\kappa = -0.2$), with $f_\mathrm{NL}$ matching the largest possible positive values allowed by CMB constraints \citep{KO09.1}.

As can be seen, due to non-Gaussianity with positive $f_\mathrm{NL}$, the power spectrum is increased at intermediate scales. At large scales the ratios get closer to unity, because the power spectra are dominated by the two-halo term, which is normalized such to reproduce the linear power spectrum, which is the same in Gaussian and non-Gaussian cosmologies. We show results for three different redshifts, indicating that the effect of non-Gaussianity is larger for higher $z$, as one might expect. The absolute increase in power at intermediate scales is quite moderate, being at most of $\sim 4\%$ at high-$z$ for the local shape and $\sim 6\%$ for the equilateral shape. The general qualitative behavior of the ratio of the non-Gaussian power spectra to the Gaussian one is quite independent of the shape of the primordial bispectrum, and also resembles the one reported by \cite{AM04.1}, that adopted completely  different non-Gaussian models. Thus, this trend seems to be a quite general property of non-Gaussian distributions with positive skewness. The fact that non-Gaussianities of the equilateral shape give a larger deviation in terms of the matter power spectrum might seem counter intuitive, since the corrections to the bias are larger in the local than in the equilateral case (\citealt{MA08.1}; \citealt*{FE09.1}). However, it should be recalled that the bounds on $f_\mathrm{NL}$ given by CMB are looser for the equilateral as compared to the local shape. As a matter of fact, we checked that considering a local non-Gaussian model with $f_\mathrm{NL} = 330$, as is the case for the equilateral model, the ratio between the non-Gaussian and the Gaussian power spectra is in fact larger than for the equilateral case, reaching up to $\sim 8\%$ for the high redshift curve.

In Figure \ref{fig:ratio} we have shown results only for positive values of $f_\mathrm{NL}$. This is because the CMB constraints on the level of non-Gaussinity, that we chose to follow in this part of the work, are highly asymmetric. For instance, the largest negative value of $f_\mathrm{NL}$ that is allowed by CMB constraints for non-Gaussianity of the local type is only $f_\mathrm{NL} = -12$, for which we expect almost no effect. As we verified, negative values of $f_\mathrm{NL}$ give rise to a symmetric behavior of the matter power spectrum around $P_\mathrm{G}(k,z)$ with respect to positive values. This is also clear from other works \citep{GI09.1} and from the following discussion in Section \ref{sct:results}.

Numerical simulations of non-Gaussian matter density fields, and computations based on renormalized perturbation theory (\citealt{GR08.2}; \citealt*{DE09.1,PI09.1,TA08.1,GI09.1}) also find an increment in the matter power spectrum due to local non-Gaussianity with positive skewness at the level of few percent. Also, they verify the qualitative behavior according to which at large scales (the only that can be probed by contemporary simulations and renormalized perturbation theory with one or two loop corrections), the effect of non-Gaussianity is milder for higher redshift. 

\begin{figure}
	\includegraphics[width=\hsize]{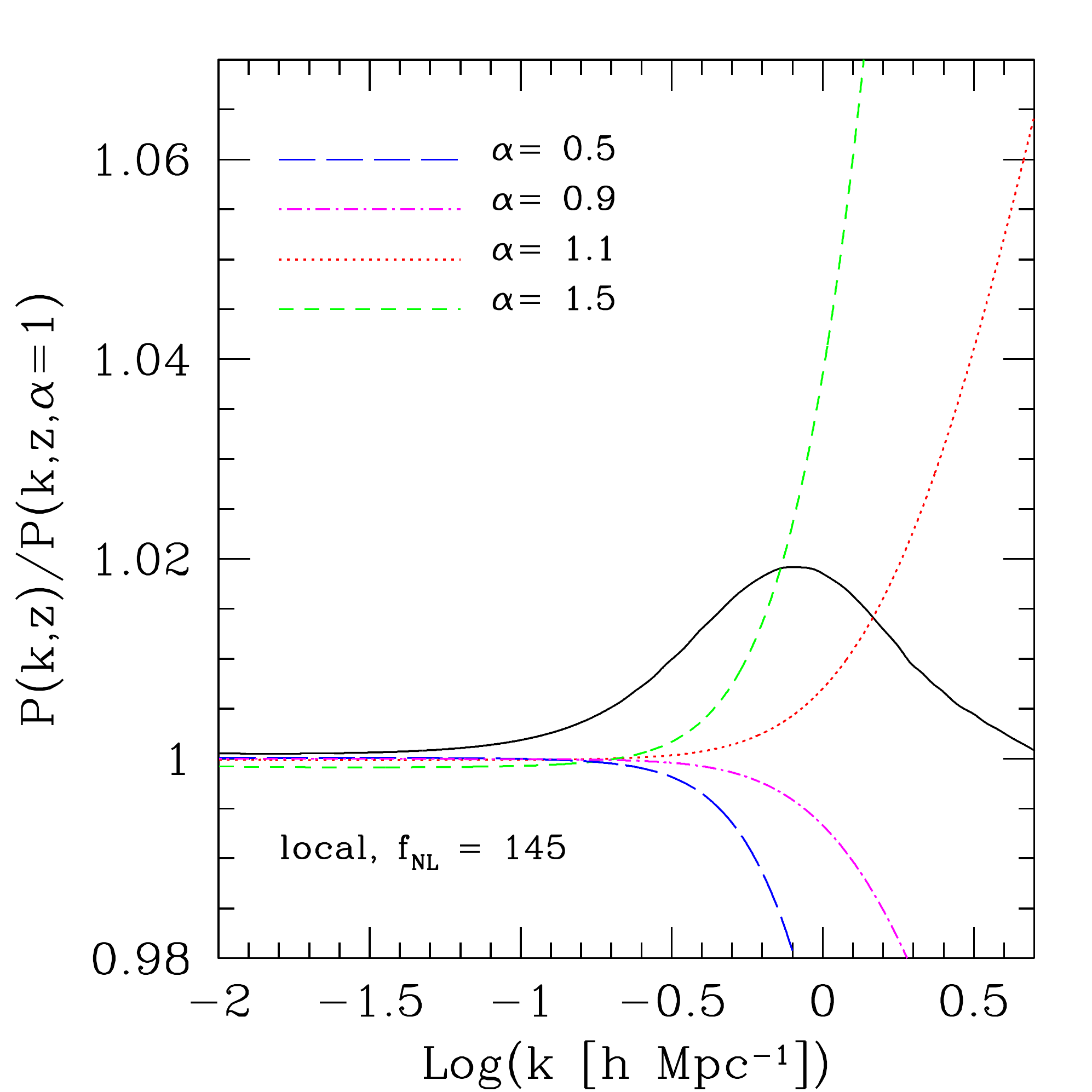}\hfill
	\caption{The ratio between the matter power spectrum in the labeled cosmology with non-Gaussian initial conditions computed assuming an inner slope for the mean dark matter halo density profile different from unity (again as labeled in the plot) to the same quantity computed instead with $\alpha=1$. For reference, we also report the ratio of the power spectrum in the non-Gaussian model to the same quantity in the Gaussian one for $\alpha=1$ (as in Figure \ref{fig:ratio}). All quantities refer to $z=0$.}
	\label{fig:ratio_alpha}
\end{figure}

There are factors that can affect the matter power spectrum differently in a non-Gaussian and Gaussian cosmology that are not captured by the simple halo model. For instance, due to the different structure formation process, it is possible that the mean halo density profile, ellipticity and substructure content are different, as well as the amount of matter outside the virial radius, all of which is neglected by the halo model. In order to exemplify one of these effects, in Figure \ref{fig:ratio_alpha} we show the effect of variations in the inner slope of average dark matter halo density profiles in non-Gaussian models, and compare it with the effect of primordial non-Gaussianity itself on the mass function and halo bias, that is instead automatically included in the halo model. As can be seen, the effect of a $\sim 10\%$ shift of the inner slope overcomes the effect of non-Gaussianity included in the halo model at scales $k \lesssim 1h$ Mpc. Larger deviations from the fiducial slope $\alpha=1$ produce even starker modifications, shifting the transition scale up to $k \sim 0.6 h$ Mpc$^{-1}$. The fact that the effect of halo density profile on the matter power spectrum can become larger than the effect of non-Gaussianity captured by the halo model was already pointed out by \citet{AM04.1}.

Another fact that is to be taken into account is that, while no secure prediction on the inner slope of dark matter halos in non-Gaussian cosmologies has been produced (with the exception of \citealt{AV03.1}, that however used very different non-Gaussian models from our own), some uncertainty on the value of $\alpha$ is also present in Gaussian cosmologies \citep{MO98.2,DI05.1}. As we verified, if the fluctuations on $\alpha$ in Gaussian and non-Gaussian cosmologies are similar, the two effects cancel, so that the ratio between the two power spectra is almost unchanged on the scales of Figure \ref{fig:ratio}. This, together with the very fact that no sharp increase (or decrease) of the matter power spectrum for non-Gaussian cosmologies with respect to the Gaussian case is detected at the scales probed by numerical simulations (e.g., \citealt{GR08.2}) indicates that there should be no large differences in the value of $\alpha$ in the two kinds of cosmological models.

In order to properly gauge the effect of primordial non-Gaussianity on the matter power spectrum that is not automatically captured by the simple halo model, a more detailed understanding of the mean dark matter halo profile, especially in non-Gaussian cosmologies, should therefore be achieved. This, as well as the assessment of the other uncertainties mentioned above, would require larger ensembles of non-Gaussian simulations and a more careful analysis thereof, which is clearly not the point in our study. Here we limit ourselves to point out that, although the general effect of non-Gaussianity is fairly captured, one should expect some higher order difference between the halo model, numerical simulations, and other ways to estimate the matter power spectrum. With this cautionary remark in mind, we proceed with the computation of the weak lensing power spectrum by setting $\alpha=1$, deferring further discussion of the effects that are not considered here to Section \ref{sct:discussion}.

%%%%%%%%%%%%%%%%%%%%%%%%%%%%%%%%%%%%%%%%%%%%%%%%%%%
\section{Results}\label{sct:results}
%%%%%%%%%%%%%%%%%%%%%%%%%%%%%%%%%%%%%%%%%%%%%%%%%%%

\begin{figure*}
	\includegraphics[width=0.45\hsize]{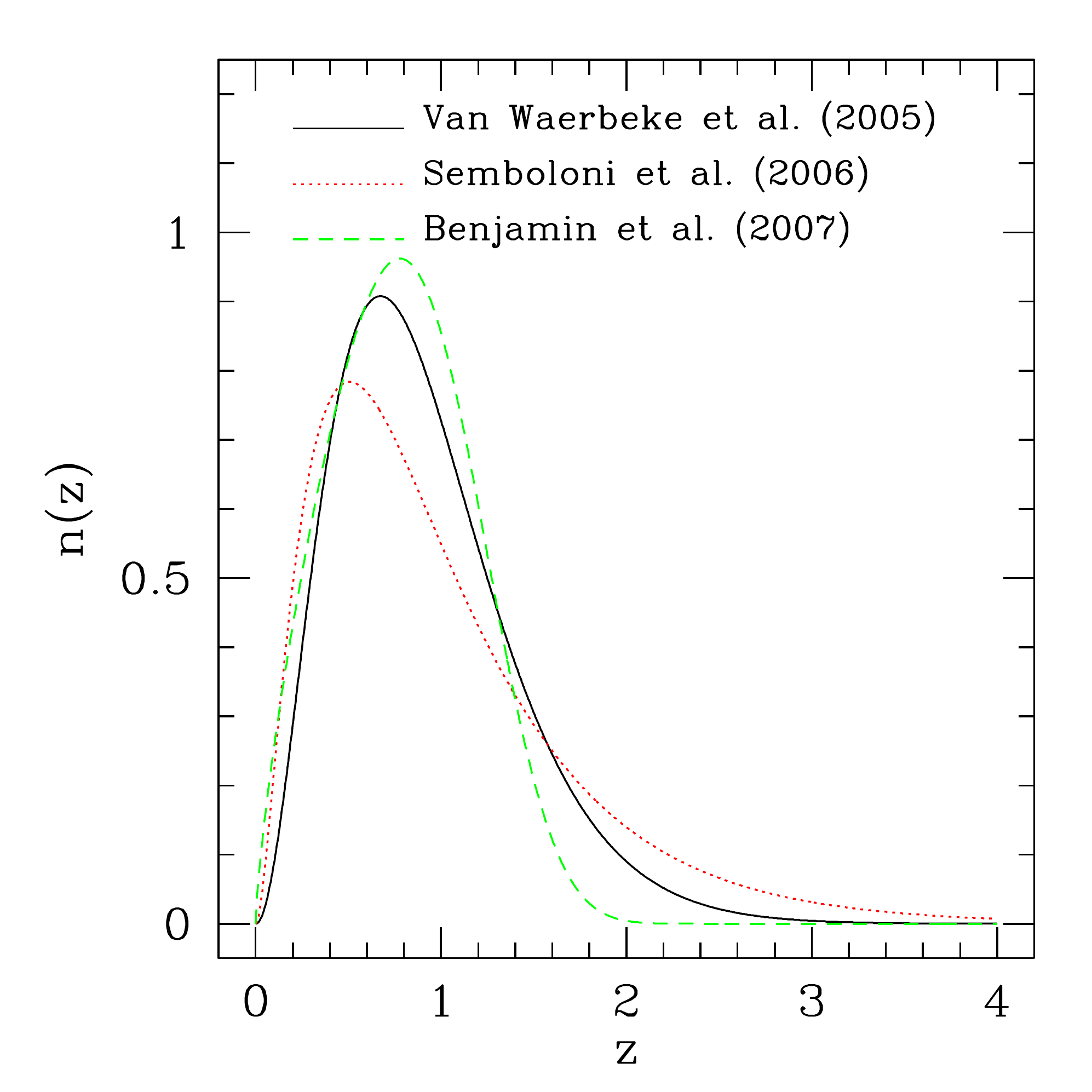}
	\includegraphics[width=0.45\hsize]{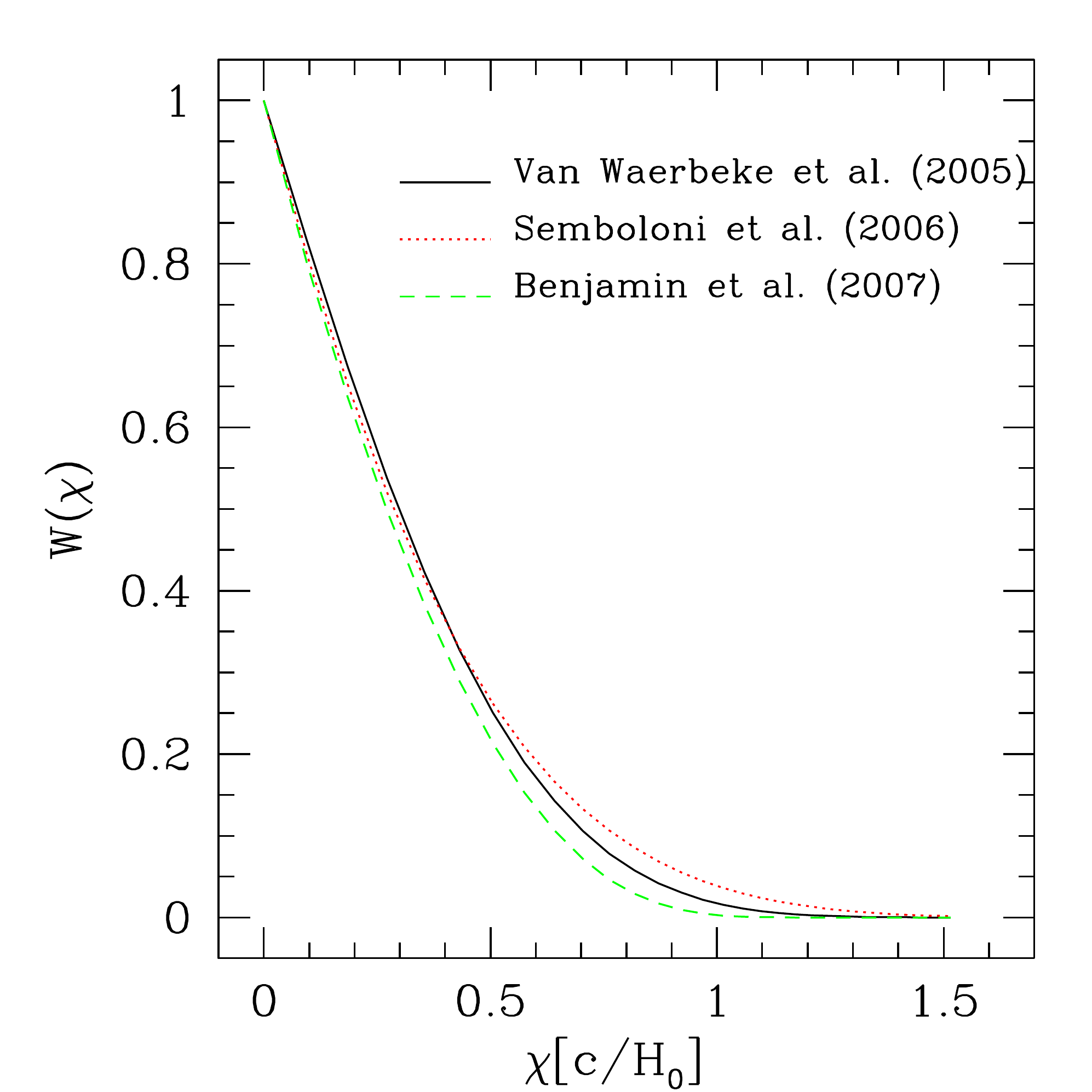}\hfill
	\caption{\emph{Left panel}. The redshift distributions fitted by \citet*{VA05.2} and \citet{SE06.1} to the HDF data and the fit obtained by \citet{BE07.2} using Eq. (\ref{eqn:z1}). \emph{Right panel}. The integration kernels corresponding to each of the redshift distributions shown on the left with the same line types and colors.} 
	\label{fig:zDist}
\end{figure*}

%%%%%%%%%%%%%%%%%%%%%%%%%%%%%%%%%%%%%%%%%%%%%%%%%%%
\subsection{Weak lensing power spectrum}
%%%%%%%%%%%%%%%%%%%%%%%%%%%%%%%%%%%%%%%%%%%%%%%%%%%

According to \citet*{BA00.1} and \citet*{BA01.1}, the power spectrum of the lensing convergence can be written as

\begin{equation}\label{eqn:wl}
C_l = \frac{9H_0^4\Omega_{\mathrm{m},0}^2}{4 c^4} \int_0^{\chi_\mathrm{H}} P\left(\frac{l}{f_K(\chi)},\chi \right) \frac{W^2(\chi)}{a^2(\chi)} d\chi,
\end{equation}
where $\chi = \chi(z)$ is the comoving distance out to redshift $z$, $a(\chi)$ is the scale factor normalized to unity today and $f_K(\chi)$ is the comoving angular-diameter distance corresponding to the comoving distance $\chi$, which depends on the spatial curvature $K$ of the Universe. The integral in Eq. (\ref{eqn:wl}) extends formally out to the horizon size $\chi_\mathrm{H}$, however the integrand becomes zero well before this limit is reached, due to the absence of sources at $z \gtrsim 4$ (see below). The redshift distribution of sources $n(z)$ has a fundamental role in the evaluation of the weak lensing power spectrum, as it defines the integration kernel

\begin{equation}
W(\chi) = \int_\chi^{\chi_\mathrm{H}} n(\chi') \frac{f_K(\chi-\chi')}{f_K(\chi')} d\chi'.
\end{equation}
The Eq. (\ref{eqn:wl}) for the convergence power spectrum was obtained using Fourier expansion and the Limber's approximation \citep{BA01.1}, while the exact expression would make use of spherical harmonic expansion. However, it was recently shown by \citet{JE09.1} that, at least when considering only the convergence power spectrum, the accuracy of the Limber's approximation is very good, better than $1\%$ at $l>10$, corresponding to $2\pi/l \lesssim 2 \times 10^3$ arcmin.

Several choices for the redshift distribution of background sources to be adopted for cosmic shear studies are available in the literature. One of the most recent ones is presented in the work of \cite{BE07.2}, where a detailed analysis of the photometric redshift distribution in four different fields is reported. The four fields considered were the Canada-France-Hawaii Telescope Legacy Survey (CFHTLS) wide survey \citep{VA02.1,HO06.1}, the GaBoDS field \citep{HE07.1}, the VIRMOS-DESCART project \citep{VA01.1,MC03.1,LE04.1}, and the RCS survey \citep{HO02.1}. \cite{BE07.2} fitted the photometric redshift distribution in the four fields using the three-parameter formula

\begin{equation}\label{eqn:z1}
n(z) = \frac{\beta}{z_0\Gamma\left[(1+\alpha)/\beta\right]} \left( \frac{z}{z_0} \right)^\alpha \exp \left[ -\left( \frac{z}{z_0} \right)^\beta \right].
\end{equation}
The same fitting formula has also been used by \citet*{VA05.2} and \cite{SE06.1} to fit the photometric redshift distribution of the Hubble Deep Field (HDF). As noted by \cite{BE07.2} and suggested by \citet{VA06.1} however, the HDF suffers of sample variance, and maybe it is also subject to a selection bias. \cite{BE07.2} noted that the formula in Eq. (\ref{eqn:z1}) does not fit very well their photometric redshift distribution if all galaxies are included, and proposed a different functional form that performed a better fit. However, when considering only their high-confidence redshift interval (outside which the fraction of catastrophic errors reaches $40-70\%$), Eq. (\ref{eqn:z1}) becomes a good fit. We chose to stick to this choice, and for each of the three parameters in Eq. (\ref{eqn:z1}) we adopted the mean of the values for the four fields analyzed by \cite{BE07.2}.

\begin{figure}
	\includegraphics[width=\hsize]{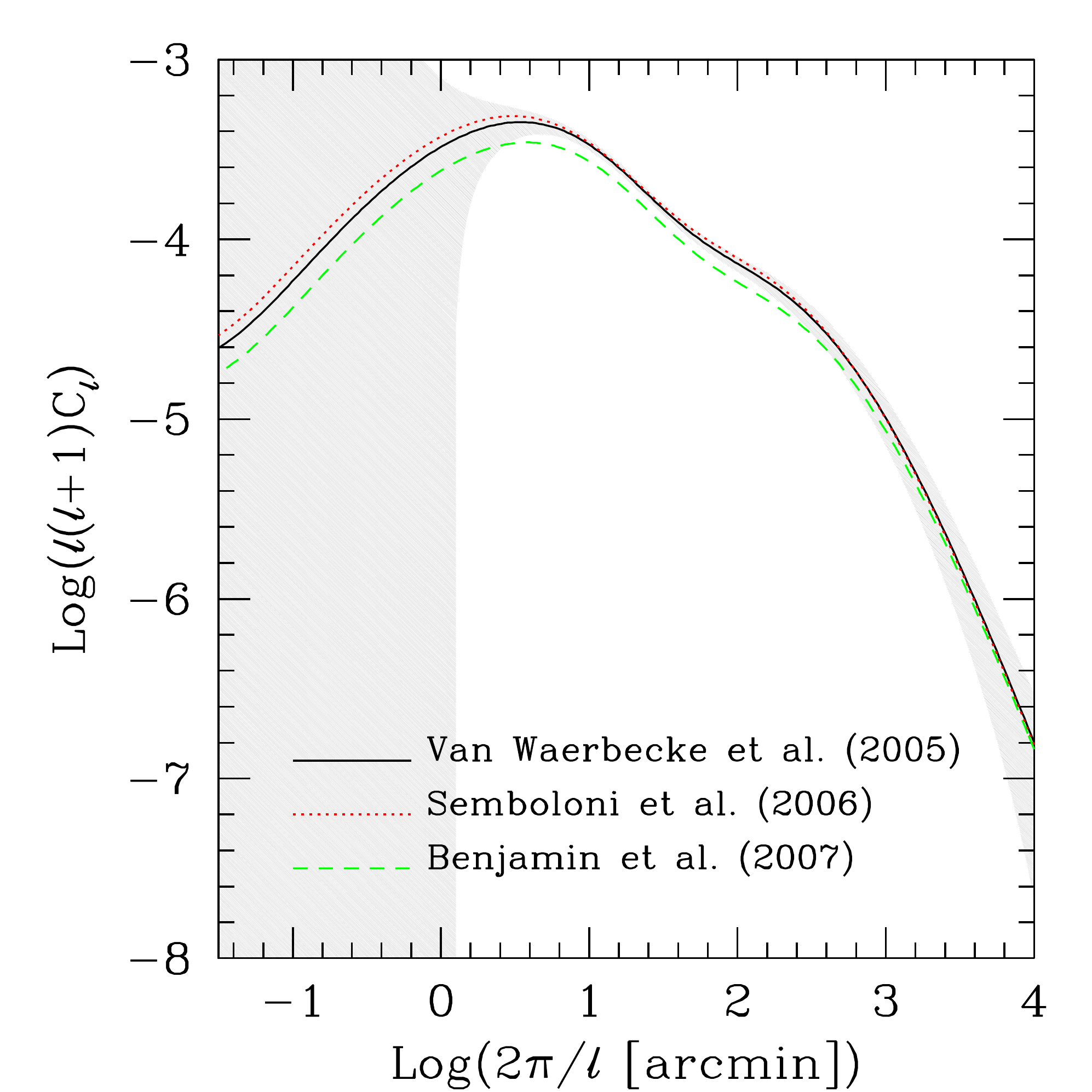}\hfill
	\caption{The fully non-linear weak lensing power spectra computed for the three different choices of source redshift distribution detailed in the text, as labelled in the plot. The gray shaded area shows statistical uncertainty on the power spectrum computed with the redshift distribution of \citet*{VA05.2}. The background cosmology is the reference Gaussian one.}
	\label{fig:wl}
\end{figure}

In order to show the effect of different choices for the background source redshift distribution, in Figure \ref{fig:zDist} we report the redshift distributions of \citet*{VA05.2}, \citet{SE06.1}, and \citet{BE07.2} with the respective integration kernels. As expected, when $z \rightarrow 0$ the kernel tends to the integral over the source redshift distribution, that is correctly normalized to unity. Independently of the assumed source redshift distribution, the kernel already vanishes at $\chi \simeq 1.5 c/H_0$, corresponding to $z \simeq 3$. In Figure \ref{fig:wl} we show the weak lensing power spectra that we computed in the $\Lambda$CDM model for the three distributions, using the matter power spectrum evaluated with the halo model described above in Section \ref{sct:modeling}. As can be seen, the difference between different power spectra can be quite significant, especially at intermediate scales, implying that the choice of the redshift distribution must be carefully addressed, given the precision level reached by future surveys.

The three power spectra tend to coincide in the linear regime at very large scales. This is because on these scales the integral in Eq. (\ref{eqn:wl}) is dominated by the low-redshift contribution, where the differences between the three integration kernels are minimal. Moving in the non-linear regime, at small scales, differences between different $n(z)$ become apparent, in particular the power spectrum for the redshift distribution of \citet{BE07.2} deviates more significantly from the power spectra for the redshift distributions of \citet*{VA05.2} and \citet{SE06.1}. This is consistent with the larger deviations that are apparent for the former distribution in the weight function $W(\chi)$. 

Also shown in the same figure is the Gaussian statistical error for the power spectrum computed for the redshift distribution of \citet*{VA05.2}. According to \cite{KA92.1,KA98.1,SE98.1,HU02.1} this has been evaluated with the prescription

\begin{equation}\label{eqn:er}
\Delta C_l = \sqrt{\frac{2}{(2l+1)\Delta lf_\mathrm{sky}}} \left( C_l + \frac{\gamma^2}{\bar{n}} \right).
\end{equation}
In Eq. (\ref{eqn:er}), $\bar{n}$ is the average number density of galaxies in the survey at hand, that we assumed equal to $\bar{n} = 40$ arcmin$^{-2}$, $f_\mathrm{sky}$ is the survey area in units of the sky area, that we posed equal to $0.5$ and $\gamma$ is the \emph{rms} intrinsic shape noise for each galaxy, that we set equal to $\gamma = 0.22$ \citep*{ZH09.1}. Such numerical values are the goals for the proposed ESA space mission EUCLID, and will be used for the rest of this paper. The parameter $\Delta l$ in Eq. (\ref{eqn:er}) represents the width of the multipole bin within which power is measured. For simplicity, we set $\Delta l = 1$ here and in the remainder of this paper. In \cite{TA07.1,TA09.1} it was shown that the multipole bin width $\Delta l$ does not significantly affect the likelihood values and the parameter estimation as long as the multipole binning is not too coarse and the weak lensing power spectrum does not show sharp variations within each bin. Finally, it has been shown that the simple Gaussian prescription in Eq. (\ref{eqn:er}) is in agreement with more elaborated error definitions on the scales where non-Gaussian errors are negligible \citep{FO08.1}.

\begin{figure*}
	\includegraphics[width=0.45\hsize]{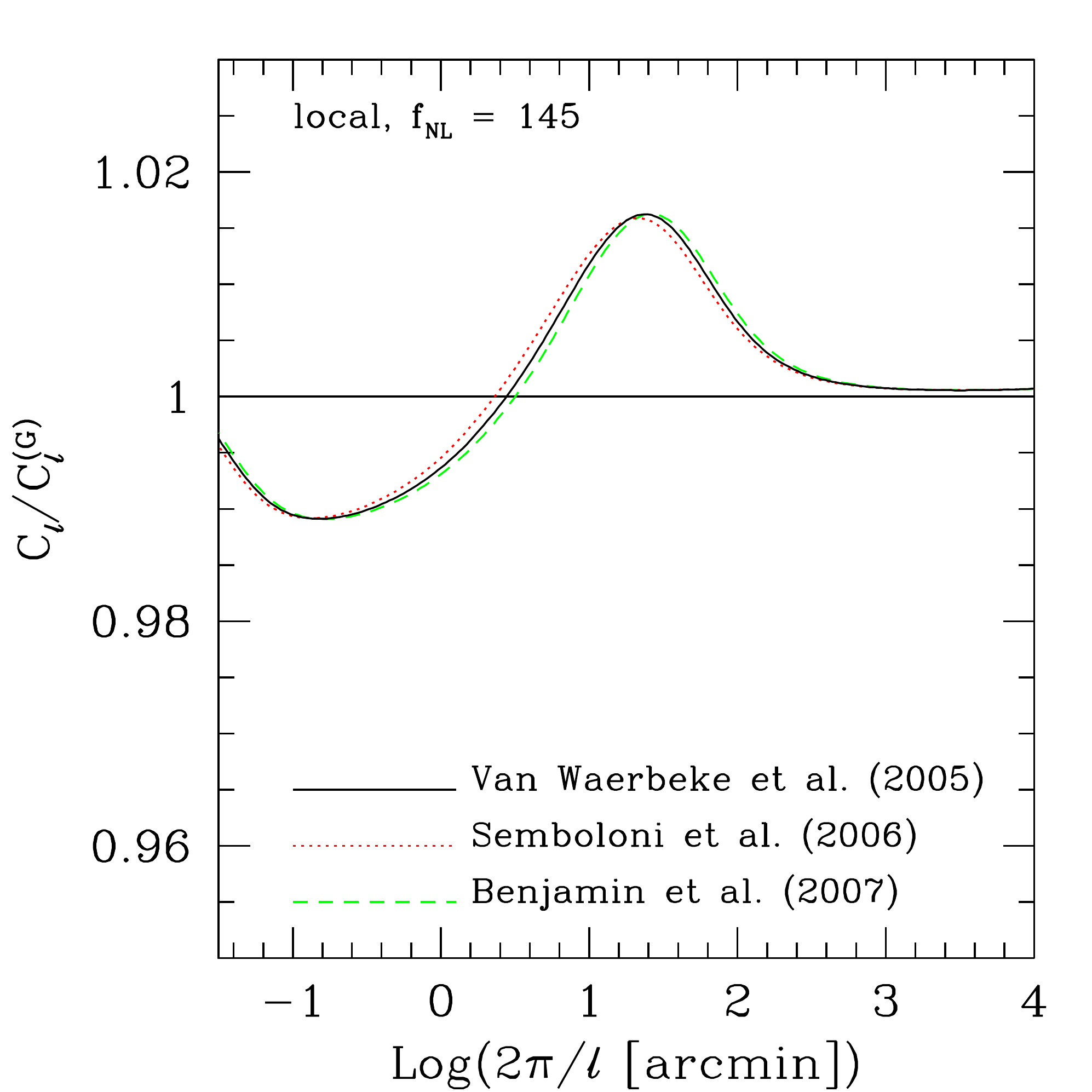}
	\includegraphics[width=0.45\hsize]{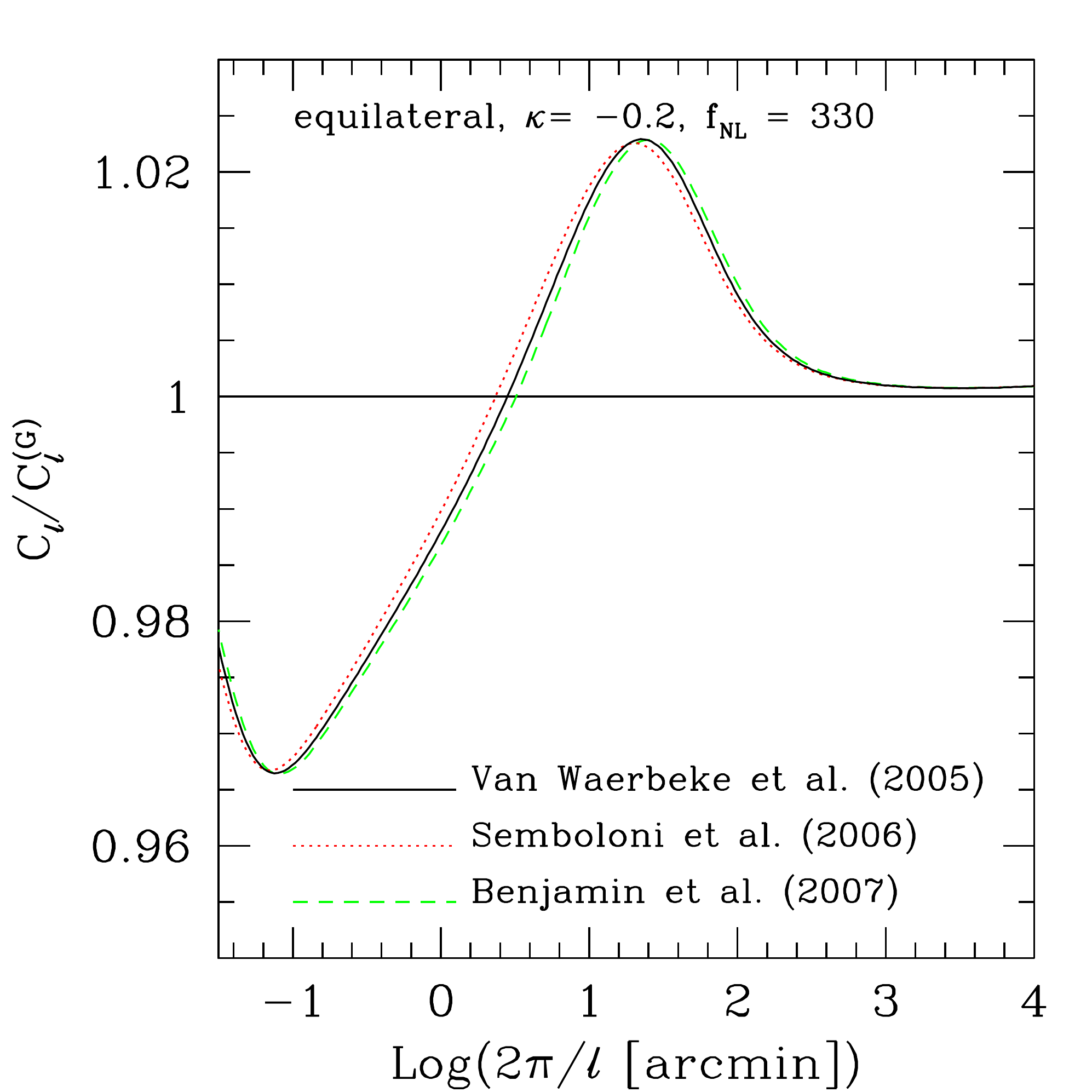}\hfill
	\caption{The ratios of weak lensing power spectra computed for the two non-Gaussian models considered here to the $\Lambda$CDM cosmology. The three source redshift distributions assumed are labeled in the plot and discussed in the text.}
	\label{fig:wlNG}
\end{figure*}

We computed the weak lensing power spectra for the non-Gaussian models with local and equilateral shapes of the primordial bispectrum, using the three source redshift distributions described above and computing the matter power spectrum as described in the previous Sections. In Figure \ref{fig:wlNG} we show the ratio of the power spectra computed in the non-Gaussian models to the reference Gaussian case. As before, the models have the maximal values of the parameter $f_\mathrm{NL}$ that are allowed by CMB constraints (see \citealt{KO09.1}).

The deviations from Gaussianity are very small, at the level of a few percent at most. Interestingly, if we exclude the small scale part of the plot, the deviation from the Gaussian model would be maximal at scales included between $\sim 10$ arcmin and $\sim 100$ arcmin, that is where the statistical error on the ratio is minimal (of the order of $\sim 10\%$ according to standard error propagation), hence we expect the bulk of the cosmological signal to come from this region. We note that the deviations from the Gaussian weak lensing power spectrum are consistent with the deviations on the three-dimensional matter power spectrum, which, e.g., in the local shape case, grow above $4\%$ only at $z>2$, where very few sources are present. The trend presented in Figure \ref{fig:wlNG}, namely of non-Gaussian power spectra being larger than the reference Gaussian one at intermediate scales and lower at very small scales is in qualitative agreement with the results of \cite{RE02.2}, implying that this might be a generic feature of all non-Gaussian models with a positive skewness. We also note that virtually no difference in this result is due to the choice of the background source redshift distribution. Therefore from this moment on, unless noted otherwise, we focused uniquely on one distribution, namely the one of \citet{BE07.2}.

%%%%%%%%%%%%%%%%%%%%%%%%%%%%%%%%%%%%%%%%%%%%%%%%%%%
\subsection{Weak lensing tomography}
%%%%%%%%%%%%%%%%%%%%%%%%%%%%%%%%%%%%%%%%%%%%%%%%%%%

\begin{figure*}
	\includegraphics[width=0.45\hsize]{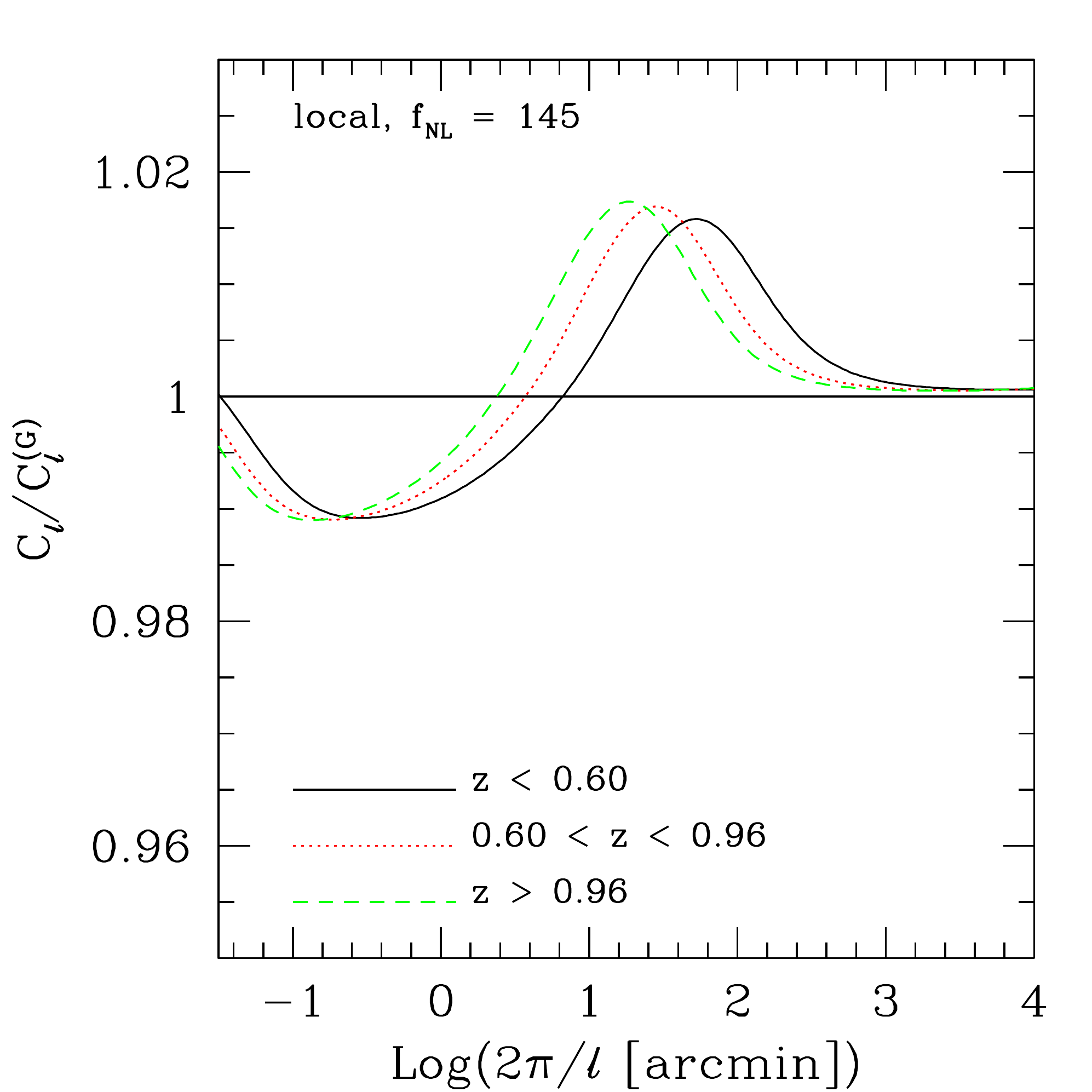}
	\includegraphics[width=0.45\hsize]{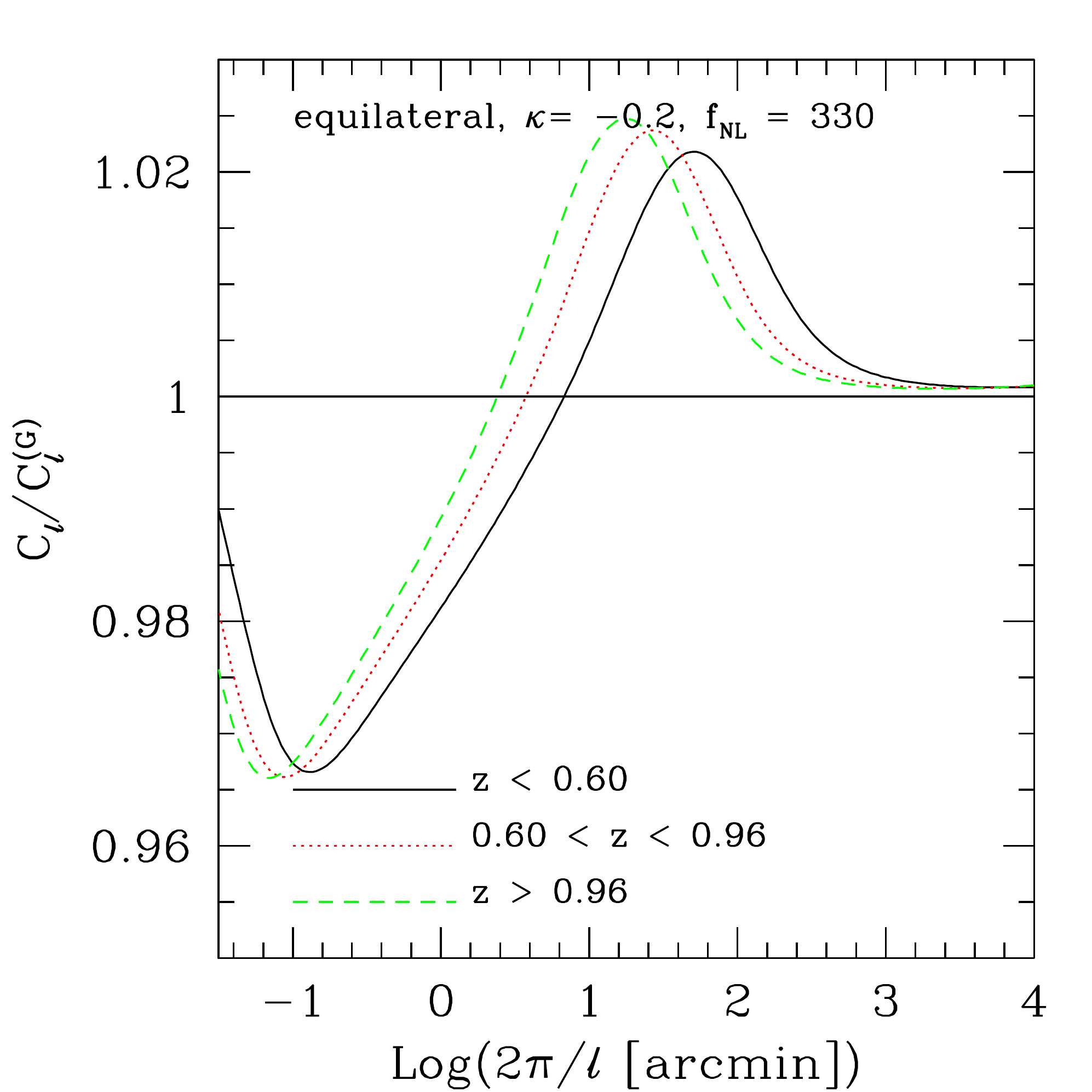}\hfill
	\caption{The ratios of the weak lensing power spectra computed in the two non-Gaussian models adopted here to the same quantity for the Gaussian cosmology. Different line types and colors refer to the three redshift bins that we adopted for weak lensing tomography, as labeled.}
	\label{fig:weakLensingZ}
\end{figure*}

It is possible to increase the amount of information that is obtainable from weak lensing surveys by employing weak lensing tomography \citep{HU99.1,TA04.1}. This consists of subdividing the redshift distribution in several bins, compute the power spectra considering those sources that are in each bin only, and then combine the information from different redshift bins. More practically, Eq. (\ref{eqn:wl}) can be generalized to consider the cross correlation power spectra for different redshift bins, as

\begin{equation}\label{eqn:wlt}
C_l^{ij} = \frac{9H_0^4\Omega_{\mathrm{m},0}^2}{4 c^4} \int_0^{\chi_\mathrm{H}} P\left(\frac{l}{f_K(\chi)},\chi \right) \frac{W_i(\chi)W_j(\chi)}{a^2(\chi)} d\chi,
\end{equation}
where the kernels now read

\begin{equation}
W_i(\chi) = \int_\chi^{\chi_\mathrm{H}} n_i(\chi') \frac{f_K(\chi-\chi')}{f_K(\chi')} d\chi'.
\end{equation}
In the previous equation, the redshift distribution $n_i$ refers to the $i-$th redshift bin, and must be normalized such that

\begin{equation}
\int_0^{\chi_\mathrm{H}}n_i(\chi)d\chi = 1
\end{equation}
for all $i$. Here, we considered three redshift bins, each one of which contains one third of the total amount of sources, adopting the distribution of \citet{BE07.2}. The two redshifts that separate the three bins are $z_1 = 0.60$ and $z_2 = 0.96$. In our case hence $i$ and $j$ run between $1$ and $n_z=3$. As noted in \citet*{MA06.1} and \citet{SU09.1}, the discriminating power of cosmic shear increases with increasing $n_z$ until $n_z = 5$, however in order to be conservative we decided not to make the redshift binning too fine, limiting ourselves to three redshift bins only.

In Figure \ref{fig:weakLensingZ} we show the ratios of the non-Gaussian power spectra computed for each of the three bins considered here to the corresponding quantities evaluated in the Gaussian cosmology. As can be seen, for higher redshift bins the peak of the deviation tends to shift toward smaller angular scales. This could be naively expected, since higher source redshift bins include higher redshift matter power spectrum information, and according to Figure \ref{fig:ratio} the peak of deviation between the Gaussian and non-Gaussian spectra shifts at smaller scales with increasing redshift. Nevertheless, no significant change in the maximum deviation is seen for the three source redshift bins considered here but, as we show below, combining information of different bins does improve the constraining power of cosmic shear. We additionally tried to recompute Figure \ref{fig:weakLensingZ} by adopting the two other source redshift distributions detailed above. We found no practical change in the effect of primordial non-Gaussianity, implying that the three $n(z)$ are too similar to each other in order for one to be appreciably preferred over the others.

We attempted a statistical analysis in order to understand what kind of constraints can be put on the value of $f_\mathrm{NL}$ by using weak lensing only, and how these improve upon inclusion of the tomographic information. We stress here that this statistical analysis is simplified, since it does not include, e.g., a proper treatment of weak lensing shear systematics and the effect of marginalization over other cosmological parameters. For this analysis, we considered multipoles included in the range between $l_1 = 50$ and $l_2 = 3000$, since little cosmological information can be extracted outside this range. Additionally, at $l>l_2$ the effect of baryon physics, that we ignored, starts to be important (\citealt{WH04.1,ZH04.1,JI06.1}), and non-Gaussian errors due to the coupling of different models caused by non-linear clustering, that we did not include, begin to be significant with respect to Gaussian errors (\citealt{WH00.1,CO01.1}).

Given all the above, the covariance matrix for weak lensing tomography has $n_z(n_z+1)n_l/2$ independent elements, where $n_l = l_2-l_1 = 2950$, and it can be written as \citep{HU06.1,MA06.1,SU09.1}

\begin{eqnarray}\label{eqn:cov}
\Gamma\left(C_l^{ij},C_{l'}^{km}\right) &=& \frac{\delta_{ll'}}{(2l+1)\Delta lf_\mathrm{sky}} \left[ \left( C_l^{ik} + \delta_{ik}\frac{\gamma^2}{\bar{n}_i} \right)\left( C_l^{jm} + \delta_{jm}\frac{\gamma^2}{\bar{n}_j} \right) \right. + 
\nonumber\\
&+& \left.\left( C_l^{im} + \delta_{im}\frac{\gamma^2}{\bar{n}_i} \right)\left( C_l^{jk} + \delta_{jk}\frac{\gamma^2}{\bar{n}_j} \right)\right]\equiv \Gamma_{\alpha\beta},
\end{eqnarray}
where $\alpha$ and $\beta$ run from $1$ through $n_z(n_z+1)n_l/2$. We note that the presence of $\delta_{ll'}$ implies that no correlation is considered between different multipoles, and this is a consequence of the fact that we ignored the non-Gaussian part of the signal covariance. This means that, if we let the matrix indices run over all the redshift bin pairs for a fixed $l$ and then change $l$ and repeat the operation, the covariance matrix (and hence its inverse) is a block diagonal matrix, where individual blocks correspond to covariance matrices between different redshift bins for a fixed multipole. Here, as before, we adopted $\Delta l = 1$ and values of the three parameters $f_\mathrm{sky}$, $\gamma$, and $\bar{n}$ corresponding to the EUCLID goals.

According to this discussion,  we can define a $\chi^2 \left(f_\mathrm{NL}\right)$ function as

\begin{equation}
\chi^2(f_\mathrm{NL}) = \sum_l \sum_{\alpha\beta}\left[\frac{}{}C_l^{\alpha} - y_l^{\alpha}(f_\mathrm{NL})\right]\Gamma^{-1}_{\alpha\beta}(l)\left[\frac{}{}C_l^{\beta} - y_l^{\beta}(f_\mathrm{NL})\right],
\end{equation}
where $\alpha$ and $\beta$ now do not run from $1$ to $n_z(n_z+1)n_l/2$, but only from $1$ through $n_z(n_z+1)/2 = 6$, i.e., the number of redshift bin independent pairs. This kind of procedure is correct as long as we can neglect the non-Gaussian part of the covariance matrix (see \citealt{TA07.1,TA09.1} for details). We assumed the measured data $C_l^{ij}$ to be the weak lensing power and cross spectra computed in the Gaussian cosmology, and the models $y_l^{ij}(f_\mathrm{NL})$ to be the spectra computed in a given non-Gaussian model with a fixed value of $f_\mathrm{NL}$. To account for the fact that the measured power and cross spectra values would not be the exact theoretical values, at each multipole we randomly perturbed the values of the spectra around the fiducial theoretical values according to a Gaussian distribution with variance given by $\Gamma\left(C_l^{ij},C_{l'}^{ij}\right)$. We repeated this procedure $128$ times, each time changing the seed for the generation of random numbers, and used the average $\chi^2(f_\mathrm{NL})$ values for the subsequent analysis.

When no tomography is applied, i.e., we have only one redshift bin, then the covariance matrix takes the form

\begin{equation}
\Gamma\left( C_l,C_{l'} \right) = \frac{\delta_{ll'}}{(2l+1)\Delta lf_\mathrm{sky}}2\left( C_l  + \frac{\gamma}{\bar{n}}\right)^2 = \delta_{ll'}\Delta C_l^2,
\end{equation}
and hence the $\chi^2(f_\mathrm{NL})$ function reads
\begin{equation}
\chi^2(f_\mathrm{NL}) =  \sum_l \frac{\left[ C_l - y_l(f_\mathrm{NL})\right]^2}{\Delta C_l^2},
\end{equation}
which is the standard $\chi^2(f_\mathrm{NL})$ definition.

\begin{table*}
  \caption{Constraints on $f_\mathrm{NL}$ at different CLs for the two cosmic shear observables considered in this work and for the local non-Gaussian model.} \label{tab:par}
  \begin{center}
    \begin{tabular}{lcccc}
      \hline
      \hline
      Probe & $68.3\%$ CL & $90\%$ CL &$95.4\%$ CL & $99\%$ CL \\
      \hline  
      Weak lensing power spectrum &$|f_\mathrm{NL}| \lesssim 17$&$|f_\mathrm{NL}| \lesssim 28$&$|f_\mathrm{NL}| \lesssim 34$&$|f_\mathrm{NL}| \lesssim 43$\\
      Weak lensing tomography &$|f_\mathrm{NL}| \lesssim 14$&$|f_\mathrm{NL}| \lesssim 23$&$|f_\mathrm{NL}| \lesssim 28$&$|f_\mathrm{NL}| \lesssim 36$\\
      \hline
      \hline
    \end{tabular}
  \end{center}
\end{table*}

Given the assumptions above, it follows that the minimum value of $\chi^2(f_\mathrm{NL})$ is reached for $f_\mathrm{NL} = 0$, and it is approximately equal to the number of degrees of freedom $\nu$. Hence, we can set $\chi^2_\mathrm{min} \equiv \chi^2(0) \simeq \nu$ and define a $\Delta \chi^2(f_\mathrm{NL})$ function as $\Delta \chi^2(f_\mathrm{NL}) \equiv \chi^2(f_\mathrm{NL})-\chi^2_\mathrm{min}$. In the simple weak lensing case we have $\nu = n_l - 1 = 2949$, while for weak lensing tomography $\nu = n_z(n_z+1)n_l/2 - 1 = 17699$. Were the non-Gaussian power and cross spectra identical to the Gaussian ones, then we would have had $\Delta \chi^2 = 0$. In any other case, $\Delta \chi^2 > 0$. The bigger $\Delta \chi^2$ is,  the better the Gaussian and non-Gaussian models can be distinguished or, in other words, at the highest confidence the non-Gaussian model can be excluded provided we measure the Gaussian power and cross spectra. For the highest value of $f_\mathrm{NL}$ consistent with CMB data in the local case, $f_\mathrm{NL} = 145$, if no tomography is applied we found $\Delta\chi^2 \simeq 73$. For the equilateral shape, for which we assumed $f_{\mathrm{NL}} = 330$, we obtained $\Delta \chi^2 \simeq 150$. These very high values of $\Delta \chi^2(f_\mathrm{NL})$ mean that these non-Gaussian models would be excluded at a very high confidence level, which would seem at odds with the results presented in Figure \ref{fig:wlNG}. As a matter of fact, we have shown in that Figure that the degree of deviation from the weak lensing power spectrum of the $\Lambda$CDM model due to primordial non-Gaussianity is much smaller than the statistical error, the former being of the order of a few percent while the latter being at least of $\sim 10\%$. However we have to remind that we are summing the signal over thousands of multipoles, hence even a very modest signal for a fixed multipole can bring to a significant integrated discriminative power. Introducing tomography and computing the $\Delta \chi^2(f_\mathrm{NL})$ as described above, the two aforementioned values raised to $\Delta \chi^2 = 104$ and $\Delta \chi^2 = 215$ respectively. The raise is expected, since adding tomographic information increases the discriminative power of the method. 

\begin{figure}
	\includegraphics[width=\hsize]{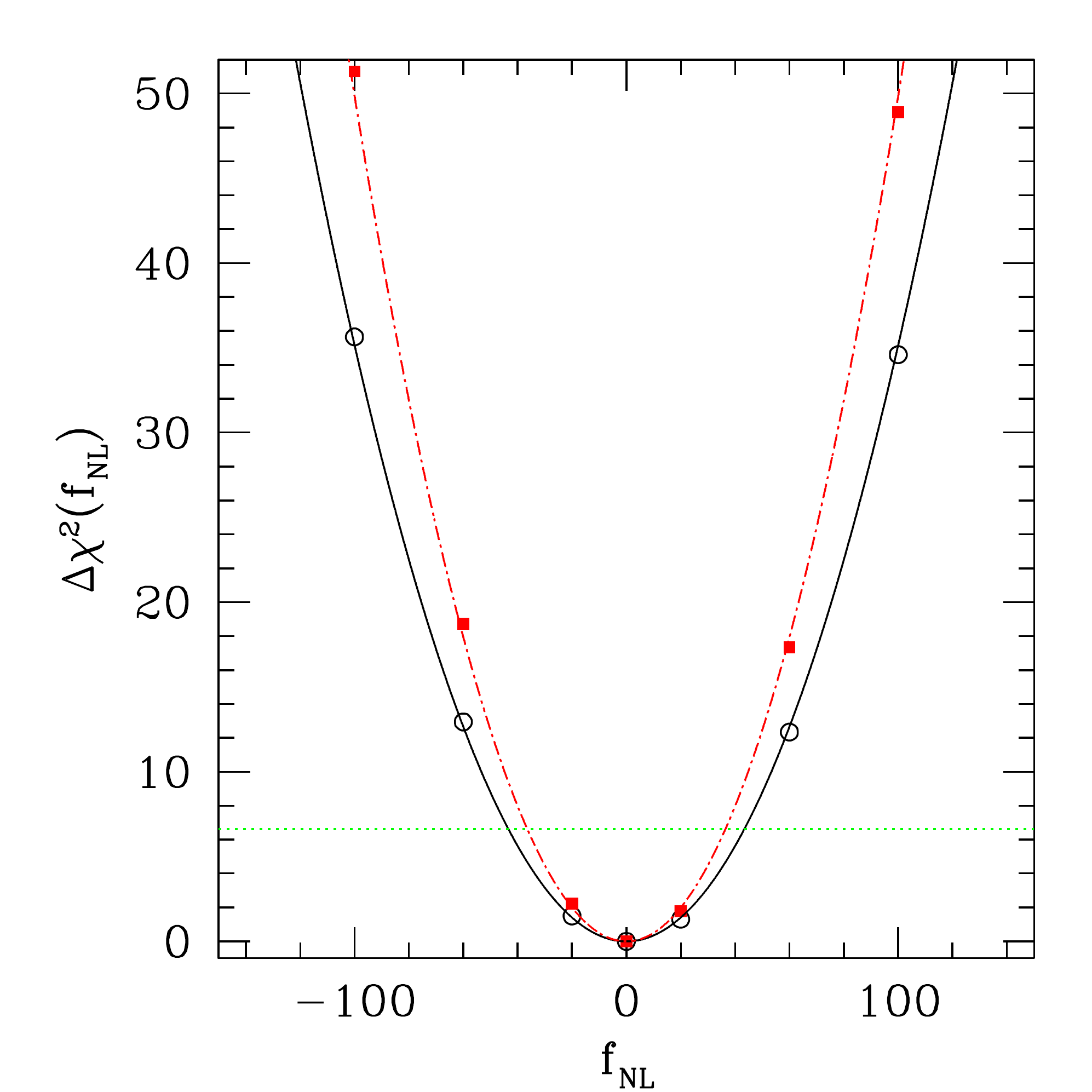}\hfill
	\caption{The $\Delta \chi^2(f_\mathrm{NL})$ function for the weak lensing power spectrum (black empty circles) and for the weak lensing tomography (red filled squares), computed assuming a non-Gaussian cosmology with local shape of the primordial bispectrum. The black solid and red dot-dashed lines represent the best fit parabolae, as detailed in the text, while the green dotted horizontal line represents the $\Delta \chi^2$ value corresponding to a $99\%$ Confidence Level detection. Note that in order to perform the fit, we also used points at $|f_\mathrm{NL}| > 100$, that are not visible in this plot.}
	\label{fig:fit}
\end{figure}

Because of the computational cost in evaluating weak lensing and cross power spectra and in order to better relate $\chi^2$ intervals into constraints on $f_\mathrm{NL}$, we computed the value of $\Delta \chi^2 (f_\mathrm{NL})$ for several different values of $f_\mathrm{NL}$, and then fitted the obtained points with some analytic expression. The result of this procedure is reported in Figure \ref{fig:fit}, where we limited ourselves to the local shape of the primordial bispectrum since it is expected to give the larger effect for a fixed $f_\mathrm{NL}$ value. We found that a parabola is an excellent fit to the $\Delta \chi^2(f_\mathrm{NL})$ function, implying the interesting fact that the power and cross spectra change linearly with the level of non-Gaussianity $f_\mathrm{NL}$. This remains true despite the slight asymmetry seen in the computed points between positive and negative $f_\mathrm{NL}$ values, which is due to the limited number of Monte-Carlo $\chi^2$ realizations that we could compute. The symmetry in the $\Delta \chi^2(f_\mathrm{NL})$ function reflects the symmetry in the behavior of the matter power spectrum with $f_\mathrm{NL}$ that we discussed in Section \ref{sct:modeling}. As one could naively expect, the $\Delta \chi^2(f_\mathrm{NL})$ function in the case of weak lensing tomography is narrower than when we consider the power spectrum only, implying that stronger constraints can be put in the former case.

Since we are operating with one single parameter and under the assumption of normally distributed errors, we can directly translate $\chi^2$ variations into Confidence Levels (CLs) for the amount of non-Gaussianity $f_\mathrm{NL}$. The kind of constraints that can be expected by a EUCLID-like survey by using weak lensing only are $|f_\mathrm{NL}| \lesssim$ few tens, with a $\sim 20\%$ improvement given by weak lensing tomography. In Table \ref{tab:par} we summarize the outcome of our statistical analysis, reporting the constraints on $f_\mathrm{NL}$ at different CLs by using both the weak lensing power spectrum only and the weak lensing tomography. As can be seen, the constraints at $99\%$ CL are already a factor of $\sim 3$ better than current WMAP constraints on positive $f_\mathrm{NL}$ values \citep{KO09.1}. At $68\%$ CL, the constraints coming from cosmic shear are expected to be competitive with other future experiments based, e.g., on the Integrated Sachs-Wolfe effect \citep*{SA67.1,CA08.1}. Currently, the only constraints coming from probes alternative to, and claimed to be more powerful than the CMB constraints are detailed in \citet{SL08.1}. There, the authors make use of Large Scale Structure data, as probed by a combination of different tracers, getting $f_\mathrm{NL}<70$ at $95\%$ CL. According to Table \ref{tab:par} our results show that the cosmic shear measurements by EUCLID should improve upon this of about a factor of $2$. In the future however, the leading constraints on the level of non-Gaussianity should still be mostly given by the CMB, with \emph{Planck} predicted to detect deviations from Gaussianity at the level of $f_\mathrm{NL} \simeq$ few \citep{SE09.1}.

%%%%%%%%%%%%%%%%%%%%%%%%%%%%%%%%%%%%%%%%%%%%%%%%%%%
\section{Discussion and conclusions}\label{sct:discussion}
%%%%%%%%%%%%%%%%%%%%%%%%%%%%%%%%%%%%%%%%%%%%%%%%%%%

In this work we computed the weak lensing power spectrum in various cosmological models with non-Gaussian initial conditions, and the relative statistical uncertainties that are expected for future large area optical surveys. The underlying non-linear matter power spectrum was evaluated using the semi-analytic halo model, that by construction depends on the internal structure of dark matter halos. As mentioned in Section \ref{sct:modeling}, it is possible that this internal structure is different in non-Gaussian with respect to Gaussian models. For instance, there has been some indication in the literature that the concentration and/or inner logarithmic slope of dark matter halos in cosmological models with primordial non-Gaussianity and positive skewness is larger than in the standard $\Lambda$CDM cosmology \citep{AV03.1}. This is intuitively in agreement with the fact that in such models it is easier to have high-density peaks, that should cross earlier the threshold for collapse, with the corresponding structures having more time to relax and compactify. A larger halo concentration would bring more power at very large wavenumbers, modifying the weak lensing power spectrum at small angular scales. On the other hand, it is not straightforward if and to what extent this expectation is fulfilled in arbitrary non-Gaussian models, and this should be verified for the models at hand with cosmological $n$-body simulations. We plan to explore this issue in the future with the numerical simulations presented in \cite{GR07.1} (see also \citealt{GR08.2}). In Any case, the effect of a different inner halo structure should show up at scales where the statistical error is very large, thus affecting our conclusions only in a minor way.

As partly discussed in the main paper body, the matter power spectrum has been estimated in numerical simulations of non-Gaussian cosmologies. However, in order to get the normalization right, the box sizes of the simulations need to be very large, implying a generically poor mass resolution. This does not allow to estimate the matter power spectrum below quasi-linear scales. A similar limitation characterizes renormalized perturbation theory, where the computation of the matter power spectrum at non-linear scales would require many loop corrections, that have not been computed yet. Semi-analytic fit to numerical simulations have been obviously calibrated in Gaussian scenarios, and it is not clear how they could be meaningfully extended to non-Gaussian models without new calibrations. Therefore, the halo model, being physically motivated, seems the only reasonable way to describe the matter power spectrum down to fully non-linear scales, despite the various limitations that have been discussed in detail in Section \ref{sct:modeling}.

In relation to this, a line of investigation that is certainly worth exploring in the future is the definition of a unique prescription for computing the matter power spectrum on all scales, that would be in agreement with simulated power spectra and the mean structure of $n$-body dark matter halos. Such prescription should also be physically motivated, in order to be straightforwardly able to comprehend the effect of baryons, dark energy and primordial non-Gaussianity. This task is going to become increasingly important, as the precision of galaxy redshift and weak lensing surveys increases.

Summarizing our results, we found that primordial non-Gaussianity has little effect on the matter power spectrum, and hence also on the cosmic shear power spectrum. This conclusion is unaffected by the choice of the background source redshift distribution, as long as the latter is observationally reasonable. Also, the shape of the primordial bispectrum seem not to have significant incidence on this qualitative conclusion. Summing the signal over a large number of multipoles can help to beat down the noise, providing a $1-$sigma detection for a level of non-Gaussianity $|f_\mathrm{NL}|\simeq 17$, if local shape for the primordial bispectrum is assumed and all other cosmological parameters are held fixed. Including weak lensing tomography can increase the constraining power of cosmic shear of $\sim 20\%$. 

These constraints are probably still looser than those that will be put with the study of e.g., number counts and correlation function of galaxy clusters in future X-ray surveys (Sartoris et al., in preparation). However, it is likely that combining these probes with cosmic shear can help breaking the degeneracy between, for instance, $f_\mathrm{NL}$ and $\sigma_8$. A more complete statistical analysis than that performed here is necessary in order to understand if and at what level this is confirmed. Finally, it is our future plan to compare the constraints given in this paper with those that can be obtained from the abundance of the S/N peaks in cosmic shear maps, a cosmological probe that has attracted some attention in the literature recently (\citealt*{BE09.1}; \citealt{DI09.1,MA09.1}). The occurrence of shear peaks depends not only on the power spectrum of large scale matter distribution, but also on the abundance of massive dark matter halos, hence it is expected to have a more constraining power with respect to cosmic shear alone.

%%%%%%%%%%%%%%%%%%%%%%%%%%%%%%%%%%%%%%%%%%%%%%%%%%%
\section*{Acknowledgments}
%%%%%%%%%%%%%%%%%%%%%%%%%%%%%%%%%%%%%%%%%%%%%%%%%%%

We acknowledge financial contributions from contracts ASI-INAF I/023/05/0, ASI-INAF I/088/06/0, and ASI "EUCLID-DUNE" I/064/08/0. We are grateful to M. Bartelmann, C. Carbone, and G. Zamorani for very useful discussions. We also wish to thank A. Amara and A. Refregier for reading the manuscript and for helpful comments. We acknowledge the anonymous referee for very useful comments that helped improving the presentation of our work.

{\small
\bibliographystyle{aa}
\bibliography{./master}
}

\end{document}